\documentclass[12pt]{amsart}

\usepackage{latexsym, amssymb, amsmath, amsthm, natbib}
\usepackage[leqno]{amsmath}
\usepackage{xy}
\xyoption{all}
\usepackage{verbatim}

\usepackage{graphicx}
\usepackage{caption,subcaption}

\usepackage[margin=1.5in]{geometry}
\usepackage{mathtools}

\usepackage{tgpagella}

\usepackage{setspace}
\usepackage{enumerate}
\usepackage[multiple]{footmisc}
\usepackage{hyperref}
\usepackage{xcolor}
\usepackage{soul}

\usepackage{aliascnt}

\usepackage{tikz}
\usetikzlibrary{positioning}

\usepackage{pgfplots}
\pgfplotsset{compat=1.14}
\usepackage{mathrsfs}
\usetikzlibrary{arrows}

\tikzset{
  every node/.style    = {
    text centered,
    line width = .5,
    anchor = center,
  },
  every label/.style   = {
    fill = white, anchor = mid,
  },
  every path/.style   = {
    > = stealth
  },
  point/.style args   = {(#1)#2}{
    rounded corners,
    fill = white,
    minimum height = 20,
    minimum width = 10,
    label = { [name = #1] above:#2 },
  },
  point a/.style args   = {(#1)#2}{
    rounded corners,
    fill = white,
    minimum height = 10,
    minimum width = 20,
    label = { [name = #1] right:#2 },
  },
  point b/.style args   = {(#1)#2}{
    rounded corners,
    fill = white,
    minimum height = 10,
    minimum width = 75,
    label = { [name = #1] right:#2 },
  },
}

\usepackage{epigraph}
\usepackage{etoolbox}
\patchcmd{\section}{\scshape}{\bfseries}{}{}
\makeatletter
\renewcommand{\@secnumfont}{\bfseries}
\makeatother

\newcommand{\cf}{C}
\newtheorem{theorem}{Theorem}

\newtheorem{definition}{Definition}
\newtheorem{lemma}{Lemma}
\newtheorem{proposition}{Proposition}
\newtheorem{claim}{Claim}

\newtheorem{defa}{\textbf Definition}

\newaliascnt{lemmaa}{thm} 
\newtheorem{lemmaa}[lemmaa]{Theorem 1'}
\aliascntresetthe{lemmaa}

\newaliascnt{propa}{thma} 
\newtheorem{propa}[propa]{Theorem 2'}
\aliascntresetthe{propa}

\theoremstyle{remark}

\newtheorem{fact}{Fact}

   \def\calx{\mathcal{X}}

\def\rational{rationalizable by} 

\newcommand{\argmax}{\mathop{\rm arg~max}\limits}

\newcommand{\ieh}[1]{{\color{magenta} IEH: #1}}
\newcommand{\fk}[1]{{\color{red} FK: #1 }}

\newcommand{\ky}[1]{{\color{violet} KY: #1 }}

\allowdisplaybreaks

\begin{document}
\setlength{\baselineskip}{20pt}
\title[Rationalizing Path-Independent Choice Rules]{Rationalizing Path-Independent Choice Rules}

\author[Yokote, Hafalir, Kojima, and Yenmez] {Koji Yokote \and Isa E. Hafalir \and Fuhito Kojima  \and M. Bumin Yenmez$^{*}$}

\thanks{
\\
An earlier version of this paper was circulated under the title ``Representation theorems for path-independent choice rules.'' We would like to thank 
David Ahn, Christopher Chambers, Federico Echenique, SangMok Lee, three anonymous referees, and the editor for their helpful comments and suggestions. Nanami Aoi, Kento Hashimoto, Asuka Hirano,
Wataru Ishida, Shinji Koiso, Sota Minowa, Daiji Nagara, Leo Nonaka, Rin Nitta, Ryosuke Sato, Kazuki Sekiya, Haruka Sawazaki, Ryo Shirakawa,  Yuji Tamakoshi, and   Kenji Utagawa provided excellent research assistance.
Koji Yokote is supported by the JSPS KAKENHI Grant-In-Aid 22KJ0717.
Fuhito Kojima is supported by the JSPS KAKENHI Grant-In-Aid 21H04979 and JST ERATO Grant Number JPMJER2301, Japan.
Yokote is affiliated with the Graduate School of Economics, the University of Tokyo, Tokyo, Japan;
Hafalir is with the UTS Business School, University of Technology Sydney, Sydney, Australia; Kojima is with the Department of Economics, the University of Tokyo, Tokyo, Japan; Yenmez is with the Department of Economics, Washington University, St. Louis, 
MO, USA and Durham University Business School, Durham, United Kingdom.
Emails: \texttt{koji.yokote@gmail.com}, \texttt{isa.hafalir@uts.edu.au}, \texttt{fuhitokojima1979@gmail.com}, \texttt{bumin@wustl.edu}.}

\begin{abstract}
Path independence is arguably one of the most important choice rule properties
in economic theory. We show that a choice rule is path independent if and only
if it is \rational{} a utility function satisfying ordinal concavity, a
concept closely related to concavity notions in discrete mathematics. We also
provide a rationalization result for choice rules that satisfy path independence and
the law of aggregate demand.
\end{abstract}

\maketitle


\section{Introduction}

\textit{Path independence} is a fundamental property of combinatorial choice rules. 
For example, consider a firm that needs to choose a subset from a set of available contracts.   
We say that the firm's choice rule is path independent if the chosen set of contracts from any set remains unchanged when 
the firm divides the set into segments, applies the rule to each segment first, and then applies the rule to the combined set of chosen contracts from all segments. This property is desirable for at least two reasons. First, without path independence, the set of chosen contracts can depend on the order in which contracts are reviewed.  
Therefore, firms whose choice rules are not path independent may not only regret rejecting offers but also may suffer from the malpractice of favoritism and other manipulations. Second, path independence is intimately related to other well-known properties of interest: A choice rule is path independent if and only if it satisfies the \textit{substitutes condition} and the \textit{irrelevance of rejected contracts} \citep{aizmal81}.

\citet{plott1973path} introduces path independence as a property of social choice formalizing
an idea in Arrow's book \textit{Social Choice and Individual Values}. 
Although originating in social choice, path independence has found applications in different areas of economic theory, 
such as market design and decision theory. \label{PIimpliesStability} 
For example, in two-sided matching markets, when agents have path-independent 
choice rules, a \textit{stable} matching exists \citep{blair88}.\footnote{The substitutes condition is not sufficient for the existence of a stable matching when choice rules are the primitive of the model rather than utility functions or preferences. 
See the discussion in \citet{aygson12a} and \citet{chayen17}.} 
\label{para:Blair}
In decision theory, path
independence and its stochastic versions have been studied extensively \citep{kalaimegiddo80}.\footnote{See also
\citet{machinaparks81} and a recent treatment by \citet{ahnechkot2018}} 
Path independence has also been studied in other fields, such as discrete mathematics, law, 
philosophy, and systems design.\footnote{For discrete mathematics see \cite{danilov2005mathematics} and  \citet{gratzer2016}, for law \citet{chapman1997} and \citet{hammond1989}, 
for philosophy \citet{rott2001} and \citet{stewart2022}, and for systems design \citet{levin1998}.}

Economic agents are usually modeled as utility maximizers: They have a utility function  
and, given a set of contracts, they choose a subset with the highest utility. 
An alternative approach is to endow agents with choice rules, for example, when agents do
not necessarily have a well-defined utility function or when the utility function is not observable.\footnote{For example,
choice rules are used to model diversity policies of schools \citep{hayeyi13,ehayeyi14,echyen12}.}
A fundamental question linking these two approaches is whether a choice rule is \textit{rationalizable} by a utility function
so that the choice from any set of contracts is the subset with the highest utility among all subsets.
Results of this nature that connect choice rules with certain properties to utility functions with corresponding properties
are called \textit{rationalization theorems}.\footnote{\label{footnote:rationalization-importance-Intro}It is, in general, useful to identify rationalization of given choice rules by preference relations or utility functions as an intellectual foundation for almost all economic analysis using preference relations or utility functions. See, for example, Chapter 1 of \citet{masco95}.} 
We provide rationalization theorems for path independent choice rules.

First, we show that a choice rule is path independent if and only if it is \rational{} a utility
function satisfying \textit{ordinal concavity} (Theorem \ref{thm1}). Ordinal concavity is a property introduced in \citet{murotashioura2003} to study discrete optimization problems.\footnote{\citet{murotashioura2003} introduce semi-strict quasi M-concavity (SSQM-concavity) as an ordinal implication of M-concavity. M-concavity is a cardinal notion, and it has a weaker variant called M$^\natural$-concavity. Analogous to weakening M-concavity to M$^\natural$-concavity, SSQM$^\natural$-concavity is the natural counterpart of SSQM-concavity (see \cite{murota2024note}). Ordinal concavity is equivalent to SSQM$^\natural$-concavity.}  
Roughly, it requires that when two sets of contracts are made closer to each other, the value of the utility function either increases on at least one side or remains unchanged on both sides. In this context, getting closer may either mean adding or removing a contract to or from the original set that we start with or the existence of a second contract such that we add one of the contracts and remove the other one.

\cite{hakoyeyo2022} show that ordinal concavity is implied by \textit{M$^{\natural}$-concavity},\footnote{See also \cite{murota2024note}.}  
which is a standard notion of concavity used in the discrete convex analysis literature. \citet{fujis03} show 
that the \emph{gross substitutes property} of \cite{kelso82} is equivalent to M$^\natural$-concavity.\footnote{M$^\natural$-concavity is also equivalent to the \textit{single-improvement property} introduced by \cite{gul99}; see Corollary 19 of \cite{murota2003new}.}     
Therefore, one implication of our result is that the difference between the gross substitutes property and the 
substitutes condition (or path independence) can be attributed to the difference between M$^\natural$-concavity 
and ordinal concavity.\footnote{To be more precise, this statement holds for rationalizable choice functions that 
satisfy the substitutes condition and for path-independent choice rules. In fact, these two classes of choice rules are the same.}

\textit{Submodularity} is another well-known condition that is often associated with a variety of substitutability notions. 
\label{para:intro-submodular}
Indeed, M$^{\natural}$-concavity, or equivalently the gross substitutes condition, implies submodularity \citep{murota2001}. Likewise, \textit{supermodularity} is used to study complementarities in different settings \citep{topkis1998}. For instance, \cite{rostekyoder2020} show that supermodularity of a utility function implies that the induced choice rule satisfies a notion of complementarities.\footnote{The setting in \cite{rostekyoder2020} is different from ours in that they allow for externalities in a multi-agent setting. Their Lemma 1 shows that when the utility function is \textit{quasisupermodular} and satisfies a single-crossing condition, the choice rule has complementarities. Quasisupermodularity is an ordinal implication of supermodularity, and the single-crossing condition is trivially satisfied when there are no externalities as in our setting.} 
Analogously, one could conjecture that submodularity implies the substitutes condition. However, submodularity and ordinal concavity are logically unrelated, and in fact rationalizability by a submodular function does not imply the substitutes condition 
or path independence.\footnote{See Claims \ref{claim:submodular} and \ref{clm:submodular-oc-independent}.} We also show that M$^\natural$-concavity and other related notions fail to give the desired rationalization result (see Section \ref{sec:relation}). 
At a high level, one of our contributions is to identify an appropriate condition on utility functions that is tightly connected with path independence and the substitutes condition (Theorem \ref{thm1}).

In the market-design literature, another choice rule property that plays a crucial role is 
\emph{the law of aggregate demand} \citep{hatmil05}. The law of aggregate demand states that 
when a larger set of contracts becomes available, the number of chosen contracts weakly increases. 
The law of aggregate demand, together with path independence, yields numerous results. 
It implies, for instance, the \textit{rural hospitals theorem} in two-sided matching markets, 
which states that the number of contracts an agent gets is the same across all stable matchings \citep{flein03}.    
In addition, in the doctor-hospital matching problem, a generalization of the doctor-proposing 
deferred-acceptance mechanism of \citet{gale62} is \textit{strategy-proof} for doctors \citep{hatmil05}.

\begin{figure}[htb]
  \begin{center}
          \includegraphics[width=14cm]{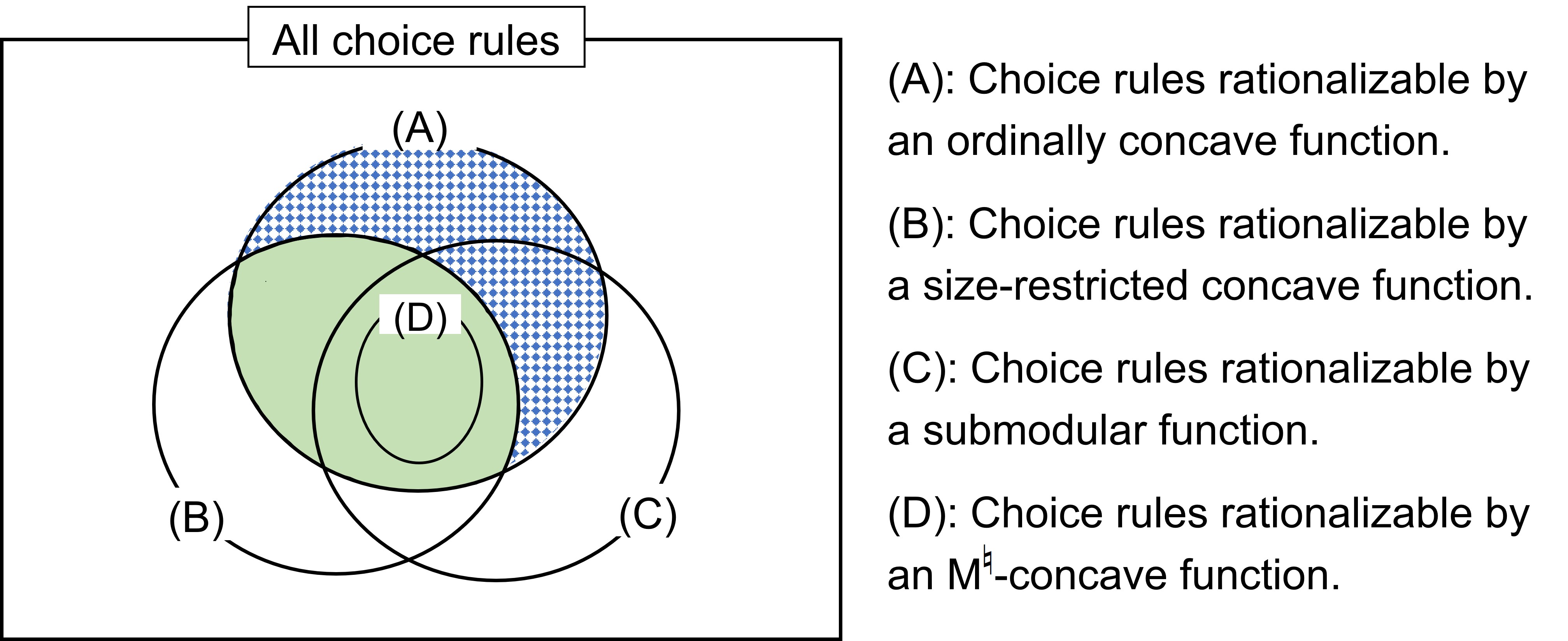}
       \end{center}
   \caption{\label{fig:ex_intro} Choice rules that are rationalizable by different notions of concavity. The union of the shaded and dotted regions, $\text{(A)}$, is the set of path-independent choice rules (Theorem \ref{thm1}). The shaded region, $\text{(A)}\cap \text{(B)}$, is the set of choice rules that satisfy path independence and the law of aggregate demand (Theorem \ref{thm2}). 
}
 \end{figure}

In our second result, we show that a choice rule satisfies path independence and the law of aggregate demand
if and only if it is \rational{} a utility function that satisfies ordinal concavity and
\textit{size-restricted concavity} (Theorem \ref{thm2}). Size-restricted concavity is a concept of discrete concavity with a quantifier such that the implication is required only for sets of
contracts with different sizes. 
Figure \ref{fig:ex_intro} illustrates the relationships between choice rules that are rationalizable by utility functions satisfying different notions of concavity.

Path independence and the law of aggregate demand are testable conditions on individual choice. \label{page:testable} Consider a data set consisting of observed choices 
of a decision maker, which may or may not include choices from all possible bundles. If there is a violation of path 
independence in the data, then we can refute the theory that the decision maker has a utility function satisfying ordinal concavity. Likewise, if there is a violation of the law of aggregate demand (even when there is no violation of path independence), then we can refute the theory that the decision maker has a utility function that satisfies ordinal 
concavity and size-restricted concavity.

The main difference between our work and the classical literature on social choice is that we assume the utility function 
(or the preference relation) is over sets of contracts, whereas in the classic setting, the utility function (or the preference relation) is over individual contracts (see \citet{moulin85} for a summary). Likewise, our choice rule
is combinatorial,\footnote{\label{footnote:combinatorial-importance}Combinatorial choice is a common characteristic of many allocation problems in the real world. It plays a crucial role in combinatorial auctions \citep{cramton2006combinatorial,de2003combinatorial}, land allocation \citep{bleakley2014land}, and allocation of airport landing slots \citep{schummer2013assignment}, among other cases. See \citet{milgrom2017discovering} who provides in-depth discussions of practical examples and highlights the importance of the combinatorial nature of many allocation problems.} 
whereas in the classical setting, multiple contracts represent the indecision of the agent, and the agent is eventually assigned only one contract.
Therefore, our results are independent of the social choice literature on rationalizability.\footnote{\label{footnote:PI-appeal}At the same time, we note that the appeal of path independence as a normatively desirable property is not affected by these differences, and hence is as relevant in our combinatorial choice context as in the classical setting. Recall our discussion in the first paragraph of the Introduction for the interpretation of path independence in our context.}

Even though path independence has been studied in the context of combinatorial choice (e.g., \cite{Echenique2007} and \cite{alva2018warp}), there are only a few papers that study rationalizability of these choice rules. 
\label{para:rationalizability}
In a recent work, \citet{yang2020} shows that path-independent choice rules are rationalizable,\footnote{\label{footnote:PI-rationalizability}For a detailed discussion of rationalizability of path-independent choice rules, see Section 5 of \citet{yang2020} as well as \cite{alva2018warp}.} but he does not provide necessary or sufficient conditions for a utility function to ensure that the corresponding choice rule satisfies path independence.
Less relatedly, \cite{chambers2018characterization} consider a setting with continuous transfers and characterize 
combinatorial demand functions that can be rationalized by quasi-linear preferences, through continuity and the law of demand.

One of the major contributions of our paper is to establish a close connection between choice rules in economics and
concavity concepts in discrete mathematics. This connection allows us to shed light on economic problems with discrete optimization techniques.
For instance, in an abstract setting, \citet{eguchi2003generalized} show that choice rules that are \rational{} M$^\natural$-concave
functions satisfy path independence and \citet{murota:metr:2013} show that they satisfy the law of aggregate demand. 
On an applied front, \citet{kojima-tamura-yokoo} build upon their results to find M$^\natural$-concave functions that rationalize a variety of practically
relevant choice rules and establish their desirable properties including computational efficiency. 
\citet{hakoyeyo2022} establish connections between ordinal concavity and choice rules in markets with dual objectives such as college admissions where
diversity and meritocracy are typical goals. While advancing this research program further, the present paper is distinctive in that it provides conditions of discrete concavity that
are \emph{equivalent} to rationalizability of desirable choice rules, thus giving a 
complete answer to a foundational issue in this research agenda.

There is also a significant literature on menu choice (where a menu is a set of items or contracts).  
\label{para:menu-choice}
In this literature, a decision maker has preferences over menus anticipating consumption of an item from 
the available menu in the future. 
In this context, \citet{kreps1979representation} provides a model of preferences for flexibility and a representation of such 
preferences using state-dependent utility functions over individual items.\footnote{Other notable papers include \cite{gul2001temptation} who provide a model of temptation and \cite{deliprus2001} who study an extension 
of \citet{kreps1979representation} that also allows preferences for commitment.} Although the decision maker in the menu choice literature ultimately consumes just one item, subsets of items or contracts arise naturally as the objects of interest as the decision maker has preferences and associated choice behavior over them, leading to a combinatorial problem such as ours.

The rest of the paper is organized as follows. We define choice rules and their properties in Section \ref{sec:prelim}, present
our rationalization results in Section \ref{sec:results}, discuss rationalizability by a utility function satisfying concavity notions other than ordinal concavity in Section \ref{sec:relation},
and conclude in Section \ref{sec:conclusion}. We provide all proofs in the Appendix.

\section{Preliminaries}\label{sec:prelim}
Let $\mathcal{X}$ denote a finite set of contracts. 
A \textbf{choice rule} is a function $\cf:2^{\mathcal{X}}\rightarrow 2^{\mathcal{X}}$ such that, for any $X\subseteq \mathcal{X}$, we have $\cf(X)\subseteq X$.\footnote{\label{footnote-contracts}In a typical model, $C$ is a choice rule of a single agent and $\mathcal{X}$ is the set of contracts that the agent can sign. We allow the agent to have multiple contracts with the same partner as in \cite{alkan03} and \cite{hatfield2012matching}. \cite{hatfield2012matching} consider a discrete setting like ours whereas \cite{alkan03} allow fractional allocations.} 
 We study two key properties of choice rules.

\begin{definition}[\citet{plott1973path}]
A choice rule $C$ satisfies \textbf{path independence} if, for any $X, X' \subseteq \calx$,
\[
C(X \cup X')=C(C(X) \cup X').
\]
\end{definition}
Path independence plays a fundamental role in social choice, market design, and decision theory.\footnote{Path independence is equivalent to the following condition: for any $X, X' \subseteq \calx$, $C(X \cup X')=C(C(X) \cup C(X'))$ \citep[Theorem 1]{plott1973path}.} It has also been used in different areas such as
discrete mathematics, law, philosophy, and systems design.\footnote{See the references in the Introduction.}


As explained in the Introduction, path independence has a normative interpretation on its own. Moreover, it is equivalent to two properties that are commonly assumed in market design. 
%
A choice rule $C$ satisfies the \textbf{substitutes condition} \citep{roth90} if, for any $X, X' \subseteq \calx$, 
\[X \supseteq X' \: \implies C(X') \supseteq  C(X) \cap X'.\]
%
%
A choice rule $C$ satisfies \textbf{irrelevance of rejected contracts} \citep{aygson12a} if, for any $X, X' \subseteq \calx$, 
\[X \supseteq X' \supseteq C(X) \: \implies C(X')=C(X).\]

\begin{proposition}[\citet{aizmal81}]\label{prop1}
A choice rule satisfies path independence if and only if it satisfies irrelevance of rejected contracts and the substitutes condition.
\end{proposition}

It is well known that a stable matching exists if the choice rule of every agent satisfies the substitutes condition and the irrelevance of rejected contracts.\footnote{\citet{roth90} show this result in a model of matching without contracts (i.e., there exists exactly one contract associated with each pair of a firm and a worker). \citet{hatmil05} generalize the result to a setting with contracts. \citet{aygson12a} point out that the result goes through only under irrelevance of rejected contracts, a condition that the proof of \citet{hatmil05} uses without explicitly assuming.} Therefore, path independence guarantees the 
existence of stable matchings in two-sided markets \citep{blair84}.

We next introduce another important property in market design called the law of aggregate demand.\footnote{\citet{alkan2002class} calls it \emph{cardinal monotonicity} and \citet{alkan03} call it \emph{size monotonicity}. These papers address matching problems without contracts, while our results hold for general matching problems with contracts.}

\begin{definition}[\citet{hatmil05}]
A choice rule C satisfies the \textbf{law of aggregate demand} if, for any $X,X' \subseteq \calx$,
\[X \supseteq X' \; \implies \; |C(X)|\geq |C(X')|.
\]
\end{definition}

The law of aggregate demand, together with path
independence, produces classic results such as the rural hospitals theorem \citep{flein03} and strategy-proofness
of a generalization of the doctor-proposing Gale-Shapley deferred-acceptance mechanism for doctors \citep{hatmil05}.

A \textbf{utility function} $u: 2^{\mathcal{X}} \rightarrow \mathbb{R}$ assigns a value to every set of
contracts.\footnote{$\mathbb{R}$ represents the set of real numbers.} A choice rule $C$ is \textbf{rationalizable}
by a utility function $u$ if, for any $X \subseteq \mathcal{X}$,
\begin{align*}
u(C(X))>u(X') \text{ for every } X' \subseteq X \text{ with } X'\neq C(X).
\end{align*}
In other words, when a choice rule is rationalizable by a utility function, from any set of available contracts, the choice rule selects
the unique subset with the highest utility.


\section{Results}\label{sec:results}
In this section, we provide two rationalization theorems using utility functions that satisfy notions of discrete 
concavity. To define these notions, we introduce some notation. 
For any $X\subseteq \mathcal{X}$ and $x\in \mathcal{X}$, let $X+x=X\cup\{x\}$ and $X-x = X \setminus \{x\}$.
Similarly, for any $X\subseteq \mathcal{X}$, let $X+\emptyset=X$ and $X-\emptyset=X$.

\begin{definition} \label{def:oc} 
A utility function $u$ satisfies \textbf{ordinal concavity} if, for any
$X, X'\subseteq \mathcal{X}$ and $x\in X\setminus X'$, there exists $x'\in (X'\setminus X)\cup \{\emptyset\}$ such that
\begin{enumerate}[(i)]
\item $u(X)<u(X-x+x')$, or
\item $u(X')<u(X'+x-x')$, or
\item $u(X)=u(X-x+x')$ and $u(X')=u(X'+x-x')$.
\end{enumerate}
\end{definition}
In other words, ordinal concavity requires that when $X$ is brought closer to  $X'$ by removing $x$ and adding $x'$, and $X'$ is brought closer to  $X$ by adding $x$ and removing $x'$,
either at least one of the two function values strictly increases or both values remain unchanged. 


Our main result is a rationalization theorem for path-independent choice rules.

\begin{theorem}\label{thm1}
A choice rule is path independent if and only if it is \rational{} a utility
function satisfying \textit{ordinal concavity}.
\end{theorem}
\begin{proof}
See Section \ref{prf:thm1}. 
\end{proof}

The if direction easily follows from the existing literature, although we provide a self-contained proof.\footnote{See 
Theorem 2 of \citet{hakoyeyo2022}.}
Our main contribution is to show the only-if direction.\footnote{In Section \ref{sec:relation}, we show that the only-if direction does {\it not} hold for other concavity notions used in prior work.} 
We note that this direction is an existence result, and its proof is constructive. 
In what follows, we give the main idea of our construction.\footnote{The proof presented in the main text is based on a suggestion made by one of the anonymous referees. We gratefully acknowledge their helpful comment. The Online Appendix provides our original proof.}

Our utility function construction relies on the decomposition lemma of \cite{aizmal81}: for any path-independent choice rule, there exists a finite sequence of linear orders $\{\succ_i\}_{i=1}^m$ over contracts such that  the choice from each set $X$ is given by the union of the most-preferred contracts according to each linear order $\succ_i$ (a formal statement of this lemma is given in Appendix \ref{sec:proofs}).\footnote{\citet{chayen17} use the decomposition lemma to make a connection between the theory of path-independent choice rules and matching theory and utilize this connection to advance both fields.} 
\label{para:utility-construction-decomposition} 
To represent each order $\succ_i$ numerically, we construct a value function $v_i$ over contracts that assigns a higher value to a more preferred contract according to $\succ_i$. We define the utility from a set $X$ by 
the weighted sum of the highest contract values among $X$ according to $v_i$ for each $i=1, \dots, m$. Then, for any $X$, the derived utility function assigns the highest utility to the chosen set $C(X)$ among all subsets of $X$, because $C(X)$ collects the most-preferred (equivalently, highest-valued) contracts according to each order $\succ_i$. Using this observation, we show that the utility function rationalizes $C$ (see step 2 in Section \ref{thm1-onlyif}).\footnote{Precisely speaking, we need to modify the utility function so that the utility from a set $X$ is smaller than that from $C(X)$ when $X\supsetneq C(X)$. See the construction of $\tilde{u}$ in equation \ref{eq:r-8}.} 
This formulation is also useful for establishing ordinal concavity because the 
highest contract value among a set is tractable when we add or remove one contract from the set (see step 3 in Section \ref{thm1-onlyif}). 

Mathematically, the construction is similar to that of \citet{kreps1979representation} in the sense that a rationalizing utility function is defined by the weighted sum of the maximum values of multiple functions. 
\label{para:Kreps-difference} We note that there are at least two differences between Kreps' model and ours. First, Kreps takes a preference over sets of contracts as primitive, while we take a choice rule as primitive. Second, Kreps assumes monotonicity of preferences in the sense that the agent prefers any set to its subsets, which cannot hold for utility functions 
rationalizing some path-independent choice rules.\footnote{For example, let $\mathcal{X}=\{x,y\}$ and, 
for any $X \subseteq \mathcal{X}$ with $x \in X$, $C(X)=\{x\}$, and, otherwise when $x \notin X$, $C(X)=\emptyset$. 
It is easy to see that this choice rule is path independent and any utility function that rationalizes it must assign a strictly higher utility to $\{x\}$ than $\{x,y\}$, violating monotonicity.} 

Theorem \ref{thm1} states that, for any path-independent choice rule, there exists {\it some} ordinally concave utility function. 
\label{para:nonuniqueness}
In general, there could be utility functions that are not ordinally concave while rationalizing a path-independent choice rule. In fact, it is obvious that the values assigned to sets of contracts that are never chosen can be lowered without changing the rationalized choice rule, and such changes in a utility function do not necessarily preserve ordinal concavity. This simple observation suggests that it is likely to be elusive to single out one condition on utility functions for the rationalization result.\footnote{\label{footnote:alkan}In fact, there is a large degree of freedom in choosing utility functions for rationalizing a given path-independent choice rule. \citet{alkan2002class} shows that the set of all chosen sets by a path-independent choice rule forms a lattice under a natural revealed preference relation. Any utility function that gives higher values to more preferred chosen subsets and gives sufficiently low values to never-chosen subsets rationalizes the choice rule.}
However, we also show that the rationalization result fails for many other conditions that are seemingly related to substitutability, so our rationalization result offers meaningful restrictions on rationalizing functions; see the discussions in the Introduction and Section \ref{sec:relation}.

In practice, it is often the case that a choice rule $C$ chooses subsets of cardinality of up to some constant $q\in \mathbb{Z}_+$, which could be interpreted as a capacity. 
\label{para:capacity-constraint}
If we consider such a choice rule in the only-if direction of Theorem \ref{thm1}, we can construct a rationalizing utility function $u$ whose effective domain  
is the subsets that have
cardinality of at most $q$.\footnote{We are grateful to an anonymous referee for their suggestion, which led us to consider this case.} 
Specifically, we can construct $u:2^{\mathcal{X}}\rightarrow \mathbb{R}\cup \{-\infty\}$ 
such that (i) $u(X)=-\infty$ for all $X\subseteq \mathcal{X}$ with $|X|>q$, (ii) $u$ rationalizes $C$, and (iii) for any $X, X'\subseteq \mathcal{X}$ with $u(X)>-\infty$ and $u(X')>-\infty$ and any $x\in X\setminus X'$, the requirement of ordinal concavity is satisfied.\footnote{To construct such a utility function, it suffices to modify the definition of $u$ in the proof (see step 1 in Section \ref{thm1-onlyif}) so that the utility from $X\subseteq \mathcal{X}$ with $|X|>q$ is equal to $-\infty$. Even though 
the stronger version of ordinal concavity that we call 
ordinal concavity$^{++}$ is no longer satisfied, 
the proof of $u$ satisfying ordinal concavity still goes through with a minor modification.} 

Next, we introduce another concavity notion that proves crucial for rationalizable choice rules to satisfy the law of aggregate demand.

\begin{definition}
A utility function $u$ satisfies \textbf{size-restricted concavity} if,
for any $X, X'\subseteq \calx$ with $|X|>|X'|$, there exists $x\in X\setminus X'$ such that
\begin{enumerate}[(i)]
\item $u(X)<u(X-x)$, or
\item $u(X')<u(X'+x)$, or
\item $u(X)=u(X-x)$ and $u(X')=u(X'+x)$.
\end{enumerate}
\end{definition}
Like ordinal concavity, this condition states that either
the function value strictly increases on at least one side or the function values remain unchanged on
both sides when two sets move closer to each other. Size-restricted concavity differs from ordinal concavity in that it requires
$X$ to have a strictly larger cardinality than $X'$, and furthermore, only one contract is added or removed when sets are brought closer to each other.
It is easy to see that size-restricted concavity is implied by M$^{\natural}$-concavity but is logically independent of ordinal concavity.\footnote{We
define M$^\natural$-concavity in Section \ref{sec:relation}. If a utility function
$u$ satisfies M$^\natural$-concavity, then it has the following property: for any $X, X'\subseteq \calx$ with $|X|>|X'|$, there exists $x\in X\setminus X'$ such that $u(X)+u(X')\leq u(X-x)+u(X'+x)$.
In fact, M$^\natural$-concavity is equivalent to this property in our setting \citep[Corollary 1.4]{murotashioura2018}.
It can be easily verified that size-restricted concavity is an ordinal implication of this inequality. Therefore, M$^\natural$-concavity
implies size-restricted concavity.}

Our second rationalization theorem uses size-restricted concavity.

\begin{theorem}\label{thm2}
A choice rule satisfies path independence and the law of aggregate demand if and only if it is
\rational{} a utility function that satisfies ordinal concavity and size-restricted concavity.
\end{theorem}
\begin{proof}
See Section \ref{prf:thm2}. 
\end{proof} 

By Theorem \ref{thm1}, a choice rule that is \rational{} an ordinally concave utility function satisfies path independence.
To complete the if direction, we show that when the utility function also satisfies size-restricted concavity, the induced choice rule satisfies the law of aggregate demand as well.
For the only-if direction, we show that the utility function we construct
in the proof of Theorem \ref{thm1} satisfies size-restricted concavity when
the choice rule satisfies both path independence and the law of aggregate demand. This completes the proof using Theorem \ref{thm1}, which shows
that the utility function satisfies ordinal concavity.

We note that the path independence of the choice rule and the ordinal concavity of the rationalizing utility function are indispensable in Theorem \ref{thm2}. More precisely, without these assumptions, the law of aggregate demand is logically unrelated to rationalizability by a size-restricted concave function.


\section{Rationalizability by concavity notions related to ordinal concavity}\label{sec:relation}

In this section, we focus on concavity notions other than ordinal concavity that have been used in economic analysis. We show that they do not constitute a necessary and sufficient condition for rationalizing choice rules with path independence  and the law of aggregate demand. 
We then discuss some variants of ordinal concavity and size-restricted concavity. 


\subsection{Submodularity}
\label{section:submodular} 
A typical assumption on functions with a combinatorial structure is 
{\it submodularity}. It is assumed in various economic models such as combinatorial auctions \citep{ausubel2002ascending} 
and cost-sharing problems \citep{moulin2001strategyproof}.   
\begin{definition} \label{def:submodularity}
A utility function $u$ satisfies \textbf{submodularity} if,
for any $X, X'\subseteq \mathcal{X}$, we have 
\begin{align*}
u(X)+u(X')\geq u(X\cup X')+u(X\cap X').
\end{align*} 
\end{definition}
Take any $Y\subseteq \mathcal{X}$ and distinct $x,y\in \mathcal{X}\setminus Y$. By substituting $Y+x$ for $X$ and $Y+y$ for $X'$ in the above definition, we obtain
\begin{align}
u(Y+x)-u(Y)\geq u(Y+x+y)-u(Y+y).
\label{eq:submodular}
\end{align}
This condition captures {\it nonincreasing marginal utility}: the marginal utility of adding $x$ to set $Y$ does not increase when $Y$ expands to $Y+y$.\footnote{It is well known that submodularity is equivalent to nonincreasing marginal utility; for a formal statement, see, e.g., Proposition 6.1 of \cite{shioura2015gross}.} In other words, if $y$ is present, then the agent is less willing to accept $x$, so this captures a kind of substitutability between $x$ and $y$. Contrary to this intuition, submodular functions do not necessarily induce substitutable or path-independent choice rules. 

\begin{claim} \label{claim:submodular} 
There exists a choice rule that is rationalizable by a submodular function but does not satisfy substitutability. In particular, the choice rule does not satisfy path independence.
\end{claim}
\begin{proof}
We provide an example in Section \ref{prf:submodular-counter}.
\end{proof} 
Therefore, Theorem \ref{thm1} does not hold if we replace ordinal concavity with submodularity. Indeed, the two notions are logically independent.
\begin{claim} \label{clm:submodular-oc-independent} 
Submodularity and ordinal concavity are logically independent of each other. 
\end{claim}
\begin{proof}
See Section \ref{prf:submodular-oc-independent}. 
\end{proof}

\subsection{M$^\natural$-concavity}
\label{section:M-concave} 
Claim \ref{claim:submodular} suggests that submodularity is too weak to capture the substitutes condition (and hence path independence). 
Therefore, we turn our attention to a stronger notion than submodularity, called {\it M$^\natural$-concavity}.  M$^\natural$-concavity is a central concavity notion in discrete convex analysis \citep{Murota:SIAM:2003}.  M$^\natural$-concavity implies submodularity \citep{murota2001}, and M$^\natural$-concavity guarantees the existence of equilibrium in markets with indivisibilities \citep{kelso82},\footnote{\citet{kelso82} use a condition called the gross substitutes condition, which is equivalent to M$^\natural$-concavity; see Section \ref{section:ordinal} for a detailed discussion.} or stable outcomes in matching markets \citep{kojima-tamura-yokoo}.

\begin{definition}
A utility function $u$ satisfies \textbf{M$^\natural$-concavity} if, for any $X, X'\subseteq \mathcal{X}$ and $x\in X\setminus X'$, there exists $x'\in (X'\setminus X)\cup \{\emptyset\}$ such that
\begin{align*}
u(X)+u(X')\leq u(X-x+x')+u(X'+x-x').
\end{align*}
\end{definition}

M$^\natural$-concavity implies ordinal concavity \citep{hakoyeyo2022}. Moreover, when a choice rule is \rational{} an
M$^\natural$-concave function, the following result holds.

\begin{proposition}[\citet{eguchi2003generalized} and \citet{murota:metr:2013}] \label{prop2}
Let $C$ be a choice rule that is \rational{} an M$^\natural$-concave utility function.
Then $C$ satisfies path independence and the law of aggregate demand.
\end{proposition}

However, M$^\natural$-concavity does not cover all choice rules that satisfy path independence and the law of aggregate demand. 
\begin{claim} \label{clm:mconcave-impossibility}
There exists a choice rule that satisfies path independence and the law of aggregate demand but is not rationalizable by any M$^\natural$-concave utility function.
\end{claim} 
\begin{proof}
We provide an example in Section \ref{prf:Mconcave-counter}.  
\end{proof}



\subsection{Ordinal content of M$^\natural$-concavity} 
\label{section:ordinal} 
M$^\natural$-concavity is a cardinal notion because the definition compares the sum of function values. 
Meanwhile, rationalization is an ordinal concept because only the ordinal ranking  
among utility values is taken into account.
Therefore,  Proposition \ref{prop2} still holds when $C$ is a choice rule that is \rational{} a monotonic transformation of
an M$^\natural$-concave utility function.\footnote{A utility function $u$ is a \textit{monotonic transformation} of another utility function $\tilde u$
if there exists a strictly increasing function $g: \mathbb{R}\rightarrow \mathbb{R}$ such that $u(X)=g(\tilde u(X))$ for all $X\subseteq \mathcal{X}$.} In other words, if we replace M$^\natural$-concavity with ``the ordinal
content of M$^\natural$-concavity'' in Proposition \ref{prop2}, the result continues to hold.\footnote{The ordinal content of cardinal
notions has drawn some attention in the literature. For example, \citet{chamech2009} study the ordinal content of supermodularity
and \citet{chamech2008} consider ordinal notions of submodularity.
}

One important implication of this discussion is that ordinal concavity is not equivalent to the ordinal content of M$^\natural$-concavity.
This can be seen by noting that under ordinal concavity, the induced choice rule does not need to satisfy the law of aggregate demand, but under the ordinal
content of M$^\natural$-concavity the law of aggregate demand holds. For example, consider a path-independent choice rule that does not
satisfy the law of aggregate demand and the corresponding utility function that we construct in the proof
of Theorem \ref{thm1}.\footnote{For example, let $\mathcal{X}=\{x,y,z\}$. Define choice rule $C$ as follows: if $x\in X$, then $C(X)=\{x\}$,
and $C(X)=X$ otherwise. It is easy to check that $C$ is path independent but does not satisfy the law of aggregate demand.}
We know that the utility function satisfies ordinal concavity by Theorem \ref{thm1}.
However, since the choice rule does not satisfy the law of
aggregate demand, the utility function cannot be a monotonic transformation of an M$^\natural$-concave utility function
 by Proposition \ref{prop2}.
In fact, even though both ordinal concavity and size-restricted concavity are implied by M$^\natural$-concavity,
their conjunction is not equivalent to its ordinal content.\footnote{If the conjunction of ordinal concavity and size-restricted concavity were equivalent to the ordinal content of M$^\natural$-concavity, then the choice rule constructed 
in the proof of  Claim \ref{clm:mconcave-impossibility}
would also be rationalizable by an M$^\natural$-concave utility function, which contradicts Claim \ref{clm:mconcave-impossibility}.}

The preceding discussion elucidates the relationships between different concepts of substitutability. The substitutes condition of \citet{hatmil05}
is closely related to the gross substitutes property of \citet{kelso82}, which is the standard concept of substitutability
in markets with continuous transfers. These two conditions are often regarded as natural counterparts in markets with and
without transfers, with the gross substitutes property being stronger than the substitutes condition.
Our analysis precisely pins down how much stronger the former is than the latter for rationalizable choice rules.
To see this, we note that \citet{fujis03} show that the gross substitutes property is equivalent to M$^\natural$-concavity in the present setting. Since
Theorem \ref{thm1} provides an equivalence result for 
 rationalizable choice rules satisfying the substitutes condition (rationalizability and the substitutes
condition are jointly equivalent to path independence, see Proposition \ref{prop1} and Theorems 4 and 5 in \cite{yang2020}), one can attribute the
difference between the two notions of substitutability to the difference between two kinds of discrete concavity, namely M$^\natural$-concavity and ordinal concavity. Similarly, Theorem \ref{thm2} shows that size-restricted concavity is precisely the additional restriction on utility functions that corresponds to imposing the law of aggregate demand on choice rules in addition to path independence.

\subsection{Pseudo M$^\natural$-concavity} \label{section:pseudoM}
\cite{hakoye2022} introduce a variant of M$^\natural$-concavity, called {\it pseudo M$^\natural$-concavity}. 
\begin{definition}
A utility function $u$ satisfies \textbf{pseudo M$^\natural$-concavity} if, for any $X, X'\subseteq \mathcal{X}$ and $x\in X\setminus X'$, there exists $x'\in (X'\setminus X)\cup \{\emptyset\}$ such that
\begin{align*}
\min\{u(X), u(X')\}\leq \min\{u(X-x+x')+u(X'+x-x')\}.
\end{align*}
\end{definition}
A notable characteristic of pseudo M$^\natural$-concavity is that upper contour sets form well-behaved discrete convex sets (see Lemma 1 of \cite{hakoye2022}). 
In view of rationalizing choice rules with path independence and the law of aggregate demand, pseudo M$^\natural$-concavity is similar to M$^\natural$-concavity: it is a sufficient condition, but not a necessary condition.


\begin{proposition} \label{prop:pseudo}
Let $C$ be a choice rule that is \rational{} a pseudo M$^\natural$-concave utility function.
Then $C$ satisfies path independence and the law of aggregate demand. 
\end{proposition}
\begin{proof}
See Section \ref{prf:pseudo}. 
\end{proof}

\begin{claim} \label{clm:pseudo-counter} 
There exists a choice rule that satisfies path independence and the law of aggregate demand but is not rationalizable by any pseudo M$^\natural$-concave utility function. 
\end{claim}
\begin{proof}
We provide an example in Section \ref{prf:pseudo-counter}.
\end{proof}  

\subsection{Variants of ordinal concavity and size-restricted concavity} \label{sect:variants} 
The preceding results reveal that Theorem \ref{thm1} does not hold if we replace ordinal concavity with submodularity, M$^\natural$-concavity, or pseudo M$^\natural$-concavity. In this section, we introduce new variants of ordinal concavity and strengthen Theorem \ref{thm1} by using these notions.

\begin{definition}\label{def:oc-}
A utility function $u$ satisfies \textbf{ordinal concavity$^-$} if, for any
$X, X'\subseteq \mathcal{X}$ and $x\in X\setminus X'$, there exists $x'\in (X'\setminus X)\cup \{\emptyset\}$ such that
\begin{enumerate}[(i)]
\item $u(X)\leq u(X-x+x')$, or
\item $u(X')\leq u(X'+x-x')$. 
\end{enumerate}
\end{definition}

To see that ordinal concavity implies ordinal concavity$^-$,\footnote{Ordinal concavity$^-$ is implied by condition \textit{QM} in \citet{murotashioura2003}.} note first that conditions (i) and (ii) of the former are the strict-inequality versions of conditions (i) and (ii) in the latter. Moreover, condition (iii) of ordinal concavity implies that both conditions (i) and (ii) of ordinal concavity$^-$ hold with equality.

\begin{definition} \label{def:oc+}
A utility function $u$ satisfies \textbf{ordinal concavity$^+$} 
if, for any $X, X'\subseteq \mathcal{X}$ and $x\in X\setminus X'$, there exists $x'\in (X'\setminus X)\cup \{\emptyset\}$ such that one of the following two conditions holds:
\begin{enumerate}[(i)]
\item $u(X)<u(X-x+x')$, or
\item $u(X')<u(X'+x-x')$. 
\end{enumerate}
\end{definition}

Ordinal concavity$^+$ implies ordinal concavity because conditions (i) and (ii) of the former are identical to conditions (i) and (ii) of the latter. We now provide a generalization of Theorem \ref{thm1} using these concavity notions.

\begin{lemmaa} \label{thm1prime}  
The following hold: 
\begin{itemize}
\item[(I)] If a choice rule is rationalizable by a utility function satisfying ordinal concavity$^-$, then it satisfies path independence. 
\item[(II)] If a choice rule is path independent, then it is rationalizable by a utility function satisfying ordinal concavity$^+$. 
\end{itemize}
\end{lemmaa}
\begin{proof}
See Section \ref{prf:thm1}.
\end{proof}

Theorem \autoref{thm1prime} implies that Theorem \ref{thm1} holds for any 
property of utility functions that implies ordinal concavity$^-$ and is implied by  
ordinal concavity$^+$. 
\label{para:OC-essential}
In particular, Theorem \ref{thm1} can be stated using either ordinal concavity$^+$ or ordinal concavity$^-$.
However, we regard rationalization by ordinal concavity as our main result for two reasons. First, ordinally concave functions have a computational advantage: there exists an algorithm that finds a maximizer of an ordinally concave function in polynomial time in the number of contracts.\footnote{See \cite{hakoyeyo2022} or \cite{murota2024note}.}
\label{para:OC-computation}
Therefore, if we can identify a utility function used in practice and verify that the function satisfies ordinal concavity, then we can compute its maximizers very fast. Second, for applications such as school choice, the utility function known to rationalize a given choice rule often satisfies ordinal concavity.\footnote{See, 
for example, the applications in \citet{kojima-tamura-yokoo}.} Ordinal concavity strikes a balance between these two properties---note that the first property does not hold if the condition on the utility function is too permissive, while the second property does not hold if the condition is too restrictive.

Furthermore, rationalization by a utility function is useful because numerical evaluations of candidates are commonplace in practice.
\label{para:OC-largeclass}
For example, a school authority often admits students based on test scores. 
The necessity part of our rationalization theorem indicates that ordinally concave
functions are a sufficiently large class of utility functions in the sense that any path-independent choice rule is ordinally concave rationalizable. Therefore, it would potentially be useful to identify computationally tractable numerical evaluations behind path-independent choice rules. 


Finally, we introduce variants of size-restricted concavity. 
\begin{definition}
A utility function $u$ satisfies \textbf{size-restricted concavity$^-$} if,
for any $X, X'\subseteq \calx$ with $|X|>|X'|$, there exists $x\in X\setminus X'$ such that
\begin{enumerate}[(i)]
\item $u(X)\leq u(X-x)$, or
\item $u(X')\leq u(X'+x)$. 
\end{enumerate}
\end{definition}

\begin{definition}
A utility function $u$ satisfies \textbf{size-restricted concavity$^+$} if,
for any $X, X'\subseteq \calx$ with $|X|>|X'|$, there exists $x\in X\setminus X'$ such that
\begin{enumerate}[(i)]
\item $u(X)<u(X-x)$, or
\item $u(X')<u(X'+x)$. 
\end{enumerate}
\end{definition}
It is easy to check that size-restricted concavity$^+$ implies size-restricted concavity. Likewise, size-restricted concavity implies size-restricted concavity$^-$. 
Similar to Theorem \autoref{thm1prime} generalizing Theorem 1, the following result generalizes Theorem \ref{thm2}.
\begin{propa} \label{thm2prime} 
The following hold: 
\begin{itemize}
\item[(I)] If a choice rule is rationalizable by a utility function satisfying ordinal concavity$^-$ and size-restricted concavity$^-$, then it satisfies path independence and the law of aggregate demand. 
\item[(II)] If a choice rule satisfies path independence and the law of aggregate demand, then it is rationalizable by a utility function satisfying ordinal concavity$^+$ and size-restricted concavity$^+$.  
\end{itemize}
\end{propa}
\begin{proof}
See Section \ref{prf:thm2}. 
\end{proof}

Therefore, Theorem \autoref{thm2prime} holds for any combination of two notions such that (i) one implies ordinal concavity$^-$ and is implied by ordinal concavity$^+$ and (ii) the other implies size-restricted concavity$^-$ and is implied by size-restricted concavity$^+$.

\section{Concluding remarks}\label{sec:conclusion}
Concavity has been a central assumption in the analysis of economies with divisible goods.
Our analysis reveals a sense in which it is essential in economies with indivisible goods as well.
In particular, we show that a path-independent choice rule is \rational{} a
utility function that satisfies a particular notion of discrete concavity, namely ordinal concavity.
In fact, we show that the relationship between path independence and rationalizability by a utility function with ordinal concavity is tight.
In economies with divisible goods, it is often the case that maximization techniques of concave functions help us characterize equilibrium outcomes.
Our rationalization theorems may prove useful in the analysis of economies with indivisible goods.

To our knowledge, the present paper is one of the first to provide rationalization theorems for combinatorial choice rules.
One possible direction for future research is to provide rationalization results for other choice rules.
It is not arduous to provide such results for canonical choice rules such as the responsive ones \citep{roth85a}
as well as the $q$-acceptant and substitutable ones \citep{kojman10}.\footnote{The rationalization results are available from the authors upon request.
They are not included in the present paper because the conditions on the utility functions are not naturally interpretable
as concavity properties, which are the main focus of the current paper.}
Meanwhile, rationalization results are still an open problem for most other choice rules, such as those with type-specific quotas \citep{abdulson03} and with reserves \citep{hayeyi13}.
More generally, it would be interesting to establish rationalization theorems for practically relevant choice rules.

\bibliographystyle{aer}
\bibliography{matching}

\appendix

\section{Proofs of Theorems in Section \ref{sec:results}}\label{sec:proofs}
In the Appendix, we provide the proofs of our results. 

\subsection{Proof of Theorems \ref{thm1} and \autoref{thm1prime}}\label{prf:thm1} 
In this section we prove Theorem \autoref{thm1prime}. As discussed in Section \ref{sect:variants}, the if and only-if directions of Theorem \ref{thm1} follow from Theorem \autoref{thm1prime} (I) and (II), respectively. 

Before starting the proof, we refer to an existing result on path-independent choice rules. 
For a choice rule $C$, we define
\begin{align}
\mathcal{X}_C=\{x\in \mathcal{X}\mid x\in C(\{x\})\}.
\label{eq:r-1}
\end{align}
If $C$ satisfies the substitutes condition, then $x\in \mathcal{X}_C$ if and only if $x\in C(X)$ for some $X\subseteq \mathcal{X}$.  
For a linear order $\succ$ over $\mathcal{X}_C$ and $X\subseteq \mathcal{X}$, we define
\begin{align*}
\text{top}(X; \succ)=\{x\in X\cap \mathcal{X}_C \mid x\succ x' \text{ for all } x'\in X\cap \mathcal{X}_C \text{ with } x'\neq x\}. 
\end{align*}
Note that $\text{top}(X, \succ)$ is a singleton or the empty set. By using this function, we can represent a path-independent choice rule $C$, as stated below.\footnote{A formal proof of this proposition can be found in \citet{moulin85} (Theorem 5) or \citet{alva2023choice} (Theorem 11.19).}
\begin{lemma}[\citet{aizmal81}]\label{prop3} 
Let $C$ be a choice rule such that $\mathcal{X}_C\neq \emptyset$. Then, $C$ satisfies path independence if and only if there exists a finite sequence of linear orders over $\mathcal{X}_C$, $\{\succ_i\}_{i=1}^m$, such that  
\begin{align}
C(X)=\bigcup_{i=1}^m \text{\normalfont top}(X, \succ_i) \text{ for all } X\subseteq \mathcal{X}.
\label{eq:AMrationalization} 
\end{align}
\end{lemma}

\subsubsection{Proof of Theorem \autoref{thm1prime} (I)} 
\label{prf:thm1-if}
Let $C$ be a choice rule and $u$ be a utility function that rationalizes $C$ and satisfies ordinal concavity$^-$. Our goal is to prove that $C$ satisfies path independence. By Proposition \ref{prop1}, it suffices to prove that $C$ satisfies the substitutes condition and irrelevance of rejected contracts. 
One easily verifies that, if a choice rule is rationalized by some utility function, then it satisfies irrelevance of rejected contracts. In the following we prove that $C$ satisfies the substitutes condition. 

Suppose, for contradiction, that the substitutes condition fails, i.e., there exist $X\subseteq \mathcal{X}$ and distinct $x,y\in X$ such that $x\in C(X)$ and $x\notin C(X-y)$.\footnote{\label{footnote:sub}We note that the substitutes condition is equivalent to the following condition: for any $X \subseteq \calx$ and any distinct $x, y\in X$, if $x\in C(X)$, then $x\in C(X-y)$. Here, we consider the negation of this condition.} 
Consider two subsets $C(X)$, $C(X-y)$ and $x\in C(X)\setminus C(X-y)$. By ordinal concavity$^-$, there exists $x'\in \bigl(C(X-y)\setminus C(X)\bigr)\cup \{\emptyset\}$ such that 
\begin{enumerate}[(i)]
\item $u(C(X))\leq u(C(X)-x+x')$, or
\item $u(C(X-y))\leq u(C(X-y)+x-x')$. 
\end{enumerate}
If $x'\neq \emptyset$, then $x'\in C(X-y)\subseteq X-y\subseteq X$, which implies \begin{align*}
C(X)-x+x'\subseteq X.
\end{align*} 
If $x'=\emptyset$, then the above set inclusion clearly holds. 
Therefore, 
condition (i) contradicts $C(X)$ uniquely maximizing $u$ among all subsets of $X$. By $x\in X-y$, we have $C(X-y)+x-x'\subseteq X-y$. Therefore, condition (ii) contradicts $C(X-y)$ uniquely maximizing $u$ among all subsets of $X-y$. 
\qed

\subsubsection{Proof of Theorem \autoref{thm1prime} (II)} 
\label{thm1-onlyif} 
Let $\cf$ be a choice rule that satisfies path independence. By Proposition \ref{prop1}, it also satisfies irrelevance of rejected contracts and the substitutes condition.
We assume that $\mathcal{X}_C\neq \emptyset$ (recall (\ref{eq:r-1})).\footnote{If $\mathcal{X}_C=\emptyset$, then $C(X)=\emptyset$ for all $X\subseteq \mathcal{X}$ (see the sentence after (\ref{eq:r-1})). For such a choice rule, we can easily construct a utility function $u$ that rationalizes $C$ and satisfies ordinal concavity; for example, $u$ defined by $u(X)=-|X|$ satisfies the desired conditions. Therefore, we consider only choice rules $C$ with $\mathcal{X}_C\neq \emptyset$.}

We proceed in three steps.
In step 1, we construct a utility function $\tilde u$.
In step 2, we prove that $\tilde u$ rationalizes $C$. 
In step 3, we prove that $\tilde u$ satisfies ordinal concavity$^+$.

\smallskip
\noindent
\textbf{Step 1: Construction of a utility function.} 

\noindent
Set $n:=|\mathcal{X}_C|\geq 1$. 
By Lemma \ref{prop3}, there exists a sequence of linear orders over $X$, $\{\succ_i\}_{i=1}^m$, such that (\ref{eq:AMrationalization}) holds. 
Let $i\in \{1, \dots, m\}$. For each $x\in \mathcal{X}_C$, let $r(x;\succ_i)\in \{1, \dots, n\}$ denote the rank of $x$ in the order $\succ_i$.\footnote{For example, if $\mathcal{X}_C=\{x,y,z\}$ and $\succ$ is given by $x\succ y \succ z$, then $r(x; \succ)=1$, $r(y;\succ)=2$, and $r(z;\succ)=3$.}
We define $v_i:\mathcal{X}_C\rightarrow \mathbb{Z}_{\geq 0}$ by 
\begin{align*}
v_i(x)=[n+1-r(x;\succ_i)]\cdot n^{m-i} & \text{ for all } x\in \mathcal{X}_C. 
\end{align*}
Notice that $v_i(x)$ gives a higher value to a contract with a higher priority according to $\succ_i$. Therefore, for any $X\subseteq \mathcal{X}$, it holds that 
$\text{\normalfont top}(X, \succ_i)=\argmax_{x\in X\cap \mathcal{X}_C}v_i(x)$.\footnote{If $X\cap \mathcal{X}_C=\emptyset$, then this equality holds in the sense that both sides coincide with the empty set.} 

We define $u:2^{\mathcal{X}} \rightarrow \mathbb{R}$ by 
\begin{align*}
u(X)=\sum_{i=1}^m \max\{v_i(x)\mid x\in X\cap \mathcal{X}_C\} \text{ for all } X\in 2^{\mathcal{X}},
\end{align*}
where the maximum over the emptyset is defined to be $0$. Hence, $X\cap \mathcal{X}_C=\emptyset$ implies $u(X)=0$. Note that, for any $X, X'\subseteq \mathcal{X}$ with $X\supseteq X'$, we have $u(X)\geq u(X')$. 
Let $\varepsilon\in (0, \frac{1}{|\mathcal{X}|})$ and we define $\tilde{u}:2^{\mathcal{X}} \rightarrow \mathbb{R}$ by 
\begin{align}
\tilde{u}(X)=u(X)-\varepsilon \cdot |X\setminus C(X)| \text{ for all } X\in 2^{\mathcal{X}}. 
\label{eq:r-8} 
\end{align}
Note that the second term on the right-hand side is sufficiently small in the sense that 
\begin{align}
\varepsilon \cdot |X\setminus C(X)|\leq \varepsilon \cdot |\mathcal{X}|<1. 
\label{eq:r-3} 
\end{align} 
We prove a claim. 
\begin{claim} \label{clm:r-1}
Let $X, X'\subseteq \mathcal{X}$. Suppose that there exists $j\in \{1, \dots, m\}$ such that
\begin{align}
&\max\{v_i(x)\mid x\in X\cap \mathcal{X}_C\}\geq \max\{v_i(x)\mid x\in X'\cap \mathcal{X}_C\} \text{ for all } i<j, and \label{eq:r-5} \\
&\max\{v_j(x)\mid x\in X\cap \mathcal{X}_C\}>\max\{v_j(x)\mid x\in X'\cap \mathcal{X}_C\}. \label{eq:r-6} 
\end{align}
Then, $\tilde{u}(X)>\tilde{u}(X')$. 
\end{claim} 
\begin{proof}
By (\ref{eq:r-6}) and the fact that $v_j$ takes integer values, we have $\max\{v_j(x)\mid x\in X\cap \mathcal{X}_C\}\geq 1$. This means that $X\cap \mathcal{X}_C\neq \emptyset$ and 
\begin{align*}
\tilde{u}(X)=u(X)-\varepsilon \cdot |X\setminus C(X)|\geq 1-\varepsilon \cdot |X\setminus C(X)|>0, 
\end{align*}
where the last inequality follows from (\ref{eq:r-3}). Therefore, the desired claim follows if $X'\cap \mathcal{X}_C=\emptyset$ (which implies $\tilde{u}(X')\leq u(X')=0$). 
In what follows, we assume that $X'\cap \mathcal{X}_C\neq \emptyset$. 

For each $i\in \{1, \dots, m\}$, let $x_i, x'_i\in \mathcal{X}_C$ be contracts such that 
\begin{align*}
v_i(x_i)=\max\{v_i(x)\mid x\in X\cap \mathcal{X}_C\}, \: v_i(x'_i)=\max\{v_i(x)\mid x\in X'\cap \mathcal{X}_C\}. 
\end{align*}
Then, 
\begin{align*}
&\tilde{u}(X)-\tilde{u}(X') \\
&=\Bigl\{\sum_{i=1}^m v_i(x_i)-\varepsilon \cdot |X\setminus C(X)|\Bigr\}-\Bigl\{\sum_{i=1}^m v_i(x'_i)-\varepsilon \cdot |X'\setminus C(X')|\Bigr\} \\
&>\sum_{i=1}^m v_i(x_i)-\sum_{i=1}^m v_i(x'_i)-1 \\
&\geq \Bigl\{v_j(x_j)-v_j(x'_j)\Bigr\}+\sum_{i=j+1}^m \Bigl\{v_i(x_i)-v_i(x'_i)\Bigr\}-1 \\
&\geq n^{m-j}+\sum_{i=j+1}^m \Bigl\{v_i(x_i)-v_i(x'_i)\Bigr\}-1 \\
&\geq n^{m-j}-\Bigl\{(n-1)\cdot n^{m-(j+1)}+\dots+(n-1)\cdot n^0\Bigr\}-1 \\
&=\cfrac{(n-1)\cdot (n^{m-j}-1)}{n-1}-\Bigl\{(n-1)\cdot n^{m-(j+1)}+\dots+(n-1)\cdot n^0\Bigr\} \\
&=0, 
\end{align*}
where
\begin{itemize}
\item the first inequality follows from (\ref{eq:r-3}) and $\varepsilon \cdot |X'\setminus C(X')|\geq 0$,
\item the second inequality follows from (\ref{eq:r-5}),  
\item the third inequality follows from (\ref{eq:r-6}) and the definition of $v_j(\cdot)$,
\item the fourth inequality follows from $v_i(x_i)\geq n^{m-i}$ and $v_i(x'_i)\leq n\cdot n^{m-i}$ for all $i$ with $j+1\leq i\leq m$,\footnote{We remark that, if $j=m$, then $\sum_{i=j+1}^m \Bigl\{v_i(x_i)-v_i(x'_i)\Bigr\}=0$, so the left-hand side of the fourth inequality is equal to $0$, and hence, the desired inequality $\tilde{u}(X)-\tilde{u}(X')\geq 0$ follows.} and
\item the last equality follows from the formula of a finite geometric series (with first term $(n-1)$ and common ratio $n$). 
\end{itemize}
The above displayed inequality establishes the desired inequality. 
\end{proof}

\smallskip
\noindent
\textbf{Step 2: Proof of {\boldmath $\tilde{u}$} rationalizing {\boldmath $C$}.}

\noindent
Let $X\subseteq \mathcal{X}$ and $Y\subseteq X$ with $Y\neq C(X)$. Our goal is to prove that
\begin{align*}
\tilde{u}(C(X))>\tilde{u}(Y).
\end{align*}
We consider two cases. 

\smallskip
\noindent
\emph{Case 1:} 
Suppose that $C(X)\subsetneq Y$. By $Y\subseteq X$ and irrelevance of rejected contracts, we have
\begin{align*}
C(Y)=C(X).
\end{align*}
By (\ref{eq:AMrationalization}), the left-hand side is equal to $\bigcup_{i=1}^m \text{\normalfont top}(Y, \succ_i)$. 
Therefore, for any $i\in \{1, \dots, m\}$, we have $\text{\normalfont top}(Y, \succ_i)\subseteq C(X)$, which together with $C(X)\subseteq Y$ implies 
\begin{align*}
\argmax_{x\in Y\cap \mathcal{X}_C}v_i(x)=\text{\normalfont top}(Y, \succ_i)=\text{\normalfont top}(C(X), \succ_i)=\argmax_{x\in C(X)\cap \mathcal{X}_C}v_i(x). 
\end{align*}
It follows that $u(C(X))=u(Y)$. By irrelevance of rejected contracts, we have $C(C(X))=C(X)$, which implies $\varepsilon\cdot |C(X)\setminus C(C(X))|=0$. Meanwhile, $C(Y)=C(X)\subsetneq Y$ implies $\varepsilon\cdot |Y\setminus C(Y)|>0$. Therefore, we obtain 
\begin{align*}
\tilde{u}(C(X))&=u(C(X))-\varepsilon\cdot |C(X)\setminus C(C(X))| \\
&=u(C(X)) \\
&=u(Y) \\
&>u(Y)-\varepsilon\cdot |Y\setminus C(Y)| \\
&=\tilde{u}(Y), 
\end{align*}
as desired. 

\smallskip
\noindent
\emph{Case 2:} Suppose that $C(X)\nsubseteq Y$, which is equivalent to $C(X)\setminus Y\neq \emptyset$. By (\ref{eq:AMrationalization}), we have $C(X)=\bigcup_{i=1}^m \text{\normalfont top}(X, \succ_i)$. Therefore, 
\begin{align}
\text{there exists $j\in \{1,\dots, m\}$ such that $\text{\normalfont top}(X, \succ_j)\in C(X)\setminus Y$.} 
\label{eq:r-7}
\end{align}
Take such a $j$ with the lowest index, i.e., $i<j$ implies $\text{\normalfont top}(X, \succ_i)\in C(X)\cap Y$. 
By $C(X)\subseteq X$ and $Y\subseteq X$, we have
\begin{align*}
\argmax_{x\in C(X)\cap \mathcal{X}_C}v_i(x)=\text{\normalfont top}(C(X), \succ_i)=\text{\normalfont top}(Y, \succ_i)=\argmax_{x\in Y\cap \mathcal{X}_C}v_i(x) \text{ for all } i<j, 
\end{align*}
which implies
\begin{align*}
\max_{x\in C(X)\cap \mathcal{X}_C}v_i(x)=\max_{x\in Y\cap \mathcal{X}_C}v_i(x) \text{ for all } i<j. 
\end{align*}
Furthermore, by (\ref{eq:r-7}) and $Y\subseteq X$, 
\begin{align*}
\max_{x\in C(X)\cap \mathcal{X}_C}v_j(x)>\max_{x\in Y\cap \mathcal{X}_C}v_j(x). 
\end{align*}
By Claim \ref{clm:r-1} applied to $C(X)$ and $Y$, we obtain the desired inequality. 

\smallskip
\noindent
\textbf{Step 3: Proof of ordinal concavity$^+$ of {\boldmath $\tilde u$}.} 
\label{para:only-if-step3}


Instead of directly proving that $\tilde u$ satisfies ordinal concavity$^+$, we show that it satisfies a stronger notion defined below. 
\begin{definition} \label{def:oc++}
A utility function $u$ satisfies \textbf{ordinal concavity$^{++}$} if, for any $X, X'\subseteq \mathcal{X}$ and $x\in X\setminus X'$, one of the following two conditions holds:
\begin{enumerate}[(i)]
\item there exists $x'\in (X'\setminus X)\cup \{\emptyset\}$ such that $u(X)<u(X-x+x')$, or
\item $u(X')<u(X'+x)$.
\end{enumerate}
\end{definition}
Note that ordinal concavity$^{++}$ implies ordinal concavity$^+$.\footnote{\label{footnote:oc++}To verify this claim, take any $X, X'\subseteq \mathcal{X}$ and $x\in X\setminus X'$. If condition (i) of ordinal concavity$^{++}$ holds for $x'\in (X'\setminus X)\cup\{\emptyset\}$, then condition (i) of ordinal concavity$^+$ holds for the $x'$. If condition (ii) of ordinal concavity$^{++}$ holds, then condition (ii) of ordinal concavity$^+$ holds for $x'=\emptyset$. Since conditions (i) and (ii) of ordinal concavity$^+$ are equivalent to the first two conditions of ordinal concavity, the above argument also shows that ordinal concavity$^{++}$ implies ordinal concavity. }



Let $X, X'\subseteq \mathcal{X}$ and $x\in X\setminus X'$. We show that 
$\tilde u$ defined by (\ref{eq:r-8}) satisfies (i) or (ii) in Definition \ref{def:oc+}.
We consider two cases. 

\smallskip
\noindent
\emph{Case 1:} Suppose that $x\notin C(X)$. By (\ref{eq:AMrationalization}), for any $i\in \{1, \dots, m\}$, we have $x\notin \text{\normalfont top}(X, \succ_i)=\argmax_{x\in X\cap \mathcal{X}_C}v_i(x)$. This condition implies 
\begin{align*}
u(X)=u(X-x). 
\end{align*}
By $x\notin C(X)$ and irrelevance of rejected contracts, we have $C(X-x)=C(X)$, which implies $|X\setminus C(X)|>|(X-x)\setminus C(X-x)|$. This inequality together with the above displayed equality implies that $\tilde{u}(X-x)>\tilde{u}(X)$. Therefore, condition (i) of ordinal concavity$^{++}$ holds for $x'=\emptyset$. 

\smallskip
\noindent
\emph{Case 2:} Suppose that $x\in C(X)$, which implies $x\in \mathcal{X}_C$ (recall the sentence after (\ref{eq:r-1})). By (\ref{eq:AMrationalization}), there exists $j\in \{1, \dots, m\}$ such that $\{x\}=\text{\normalfont top}(X, \succ_j)$. Take such a $j$ with the lowest index, i.e., 
\begin{align}
\text{$i<j$ implies $\{x\}\neq \text{\normalfont top}(X, \succ_j)$.}
\label{eq:r-9} 
\end{align}

\smallskip
\noindent
\emph{Subcase 2-1:} Suppose that 
\begin{align}
\text{there exists $x'\in X'\cap \mathcal{X}_C$ such that $v_j(x')>v_j(x)$.} 
\label{eq:r-10}
\end{align}
Since $\{x\}=\text{\normalfont top}(X, \succ_j)=\argmax_{y\in X\cap \mathcal{X}}v_j(x)$, we have $x'\in (X'\setminus X) \cap \mathcal{X}$. We also have 
\begin{align*}
&\max_{y\in (X-x+x')\cap \mathcal{X}_C} v_i(y)\geq \max_{x\in X\cap \mathcal{X}_C} v_i(y) \text{ for all } i<j, \text{ and } \\
&\max_{y\in (X-x+x')\cap \mathcal{X}_C} v_j(y)>\max_{y\in X\cap \mathcal{X}_C} v_j(y). 
\end{align*}
where the former inequality follows from (\ref{eq:r-9}) and the latter inequality follows from $\{x\}=\text{\normalfont top}(X, \succ_j)=\argmax_{x\in X\cap \mathcal{X}_C} v_j(x)$ and (\ref{eq:r-10}). Claim \ref{clm:r-1} applied to $X-x+x'$ and $X$ implies that condition (i) of ordinal concavity$^{++}$ holds. 

\smallskip
\noindent
\emph{Subcase 2-2:} Suppose that $v_j(x)>v_j(x')$ for all $x'\in X'\cap \mathcal{X}_C$.\footnote{This case subsumes $X'\cap \mathcal{X}_C=\emptyset$.}  
Then,
\begin{align*}
\max_{y\in (X'+x)\cap \mathcal{X}_C}v_j(y)=v_j(x)>\max_{y\in X'\cap \mathcal{X}_C}v_j(y). 
\end{align*}
Since $v_j(\cdot)$ takes only integer values, the above inequality implies that the left-hand side is bigger than the right-hand side by no less than 1. Therefore, 
\begin{align}
u(X'+x)\geq u(X')+1. 
\label{eq:r-11} 
\end{align}
It follows that
\begin{align*}
&\tilde{u}(X'+x)-\tilde{u}(X') \\
&=\bigl\{u(X'+x)-\varepsilon \cdot |(X'+x)\setminus C(X'+x)|\bigr\}-\bigl\{u(X')-\varepsilon \cdot |X'\setminus C(X')|\bigr\} \\
&\geq u(X'+x)-\varepsilon \cdot |(X'+x)\setminus C(X'+x)|-u(X') \\
&\geq 1-\varepsilon \cdot |(X'+x)\setminus C(X'+x)| \\
&>0, 
\end{align*}
where 
\begin{itemize}
\item the first inequality follows from $\varepsilon\cdot |X'\setminus C(X')|\geq 0$, 
\item the second inequality follows from (\ref{eq:r-11}), and
\item the last inequality follows from (\ref{eq:r-3}).
\end{itemize}
We conclude that condition (ii) of ordinal concavity$^{++}$ holds. As ordinal concavity$^{++}$ implies ordinal concavity$^{+}$, the desired conclusion follows.
\qed

\subsection{Proof of Theorems \ref{thm2} and \autoref{thm2prime}} \label{prf:thm2}
In this section, we prove Theorem \autoref{thm2prime}. Note that the if and only-if directions of Theorem \ref{thm2} follow from Theorem \autoref{thm2prime} (I) and (II), respectively. 

\subsubsection{Proof of Theorem \autoref{thm2prime} (I)} \label{prf:thm2-if} 
Let $\cf$ be a choice rule and $u$ be a utility function that rationalizes $\cf$ and satisfies ordinal concavity$^-$ and size-restricted concavity$^-$.
By Theorem \autoref{thm1prime} (I), $C$ satisfies path independence. It remains to prove that $C$ satisfies the law of aggregate demand.
The proof is similar to that of Theorem 3.10 in \citet{murota:dca:2016}. Suppose, for contradiction, that the law of aggregate demand is violated, i.e., there exist $X$ and $X'$ such that $X\supseteq X'$ and $|C(X')|>|C(X)|$. By size-restricted concavity$^-$ of $u$, there exists $x\in C(X')\setminus C(X)$ such that
\begin{enumerate}[(i)]
\item $u(C(X'))\leq u(C(X')-x)$, or
\item $u(C(X))\leq u(C(X)+x)$. 
\end{enumerate}
If (i) holds, then we obtain a contradiction to $C(X')$ uniquely maximizing $u$ among all subsets of $X'$.
If (ii) holds, then together with $x\in C(X')\subseteq X'\subseteq X$, we obtain a contradiction to $C(X)$ uniquely maximizing $u$ among all subsets of $X$. 

\subsubsection{Proof of Theorem \autoref{thm2prime} (II)}
Let $\cf$ be a choice rule that satisfies path independence and the law of aggregate demand.
We define $\tilde u$ as in (\ref{eq:r-8}).
As proven in Section \ref{thm1-onlyif}, $\tilde u$ rationalizes $C$ and satisfies ordinal concavity$^+$.
It remains to prove that $\tilde u$ satisfies size-restricted concavity$^+$.
Let $X, X'\subseteq \mathcal{X}$ with $|X|>|X'|$. We consider two cases.

\smallskip
\noindent
\emph{Case 1:}\footnote{The proof of this case is similar to that of Subcase 2-2 of step 3 in Section \ref{thm1-onlyif}.} 
Suppose that there exist $x\in X\setminus X'$ and $j\in \{1, \dots, m\}$ such that 
\begin{align*}
\max_{y\in (X'+x)\cap \mathcal{X}_C}v_j(y)>\max_{y\in X'\cap \mathcal{X}_C}v_j(y). 
\end{align*}
Since $v_j(\cdot)$ takes only integer values, the left-hand side is bigger than the right-hand side by no less than 1. Therefore, 
\begin{align}
u(X'+x)\geq u(X')+1. 
\label{eq:r-12} 
\end{align}
It follows that
\begin{align*}
&\tilde{u}(X'+x)-\tilde{u}(X') \\
&=\bigl\{u(X'+x)-\varepsilon \cdot |(X'+x)\setminus C(X'+x)|\bigr\}-\bigl\{u(X')-\varepsilon \cdot |X'\setminus C(X')|\bigr\} \\
&\geq u(X'+x)-\varepsilon \cdot |(X'+x)\setminus C(X'+x)|-u(X') \\
&\geq 1-\varepsilon \cdot |(X'+x)\setminus C(X'+x)| \\
&>0, 
\end{align*}
where 
\begin{itemize}
\item the first inequality follows from $\varepsilon\cdot |X'\setminus C(X')|\geq 0$, 
\item the second inequality follows from (\ref{eq:r-12}), and
\item the last inequality follows from (\ref{eq:r-3}).
\end{itemize}
We conclude that condition (ii) of size-restricted concavity$^+$ holds.

\smallskip
\noindent
\emph{Case 2:} Suppose that, for any $x\in X\setminus X'$, we have 
\begin{align*}
\max_{y\in (X'+x)\cap \mathcal{X}_C}v_i(y)=\max_{y\in X'\cap \mathcal{X}_C}v_i(y) \text{ for all } i\in \{1, \dots, m\}.  
\end{align*}
This condition implies  
\begin{align*}
\max_{y\in (X\cup X')\cap \mathcal{X}_C}v_i(y)=\max_{y\in X'\cap \mathcal{X}_C}v_i(y) \text{ for all } i\in \{1, \dots, m\}, 
\end{align*}
which is equivalent to 
\begin{align*}
\text{\normalfont top}(X\cup X', \succ_i)=\text{\normalfont top}(X', \succ_i) \text{ for all } i\in \{1, \dots, m\}. 
\end{align*}
By (\ref{eq:AMrationalization}), we have 
\begin{align}
C(X\cup X')=C(X').
\label{eq:r-13}
\end{align} 

\smallskip
\noindent
\emph{Subcase 2-1:} \label{para:theorem2-onlyif-impossible} Suppose that $X\setminus X' \subseteq C(X)$. Since $C$ satisfies path independence, by Proposition \ref{prop1}, $C$ satisfies the substitutes condition. Hence, the following set-inclusion holds:\footnote{To see that (\ref{1-1-2}) holds, suppose that there exists $x\in C(X\cup X') \setminus C(X)$ with $x\notin X'\setminus X$.
By $x\in C(X\cup X')$ and $C(X\cup X') \subseteq X'$ (the latter condition follows from (\ref{eq:r-13})), we have $x\in X'$.
Together with $x\notin X'\setminus X$, it implies $x\in X\cap X'$.
By combining $x\notin C(X)$, $x\in X$, and $x\in C(X\cup X')$, we obtain a contradiction to the substitutes condition.}
\begin{align}
C(X\cup X') \setminus C(X)\subseteq X' \setminus X.
\label{1-1-2}
\end{align}
Since $|X|>|X'|$,
\begin{align}
|X\setminus X'|>|X'\setminus X|.
\label{1-1-3}
\end{align}
Then,
\begin{align*}
|C(X\cup X') \setminus C(X)| \leq |X'\setminus X|<|X\setminus X'|\leq |C(X)\setminus C(X\cup X')|,
\end{align*}
where the first inequality follows from (\ref{1-1-2}), the second inequality follows from (\ref{1-1-3}), and the last
inequality follows from $X\setminus X'\subseteq C(X) \setminus X' \subseteq C(X)\backslash C(X\cup X')$, where
\begin{itemize}
\item the first set-inclusion follows from the assumption of Subcase 2-1, and
\item the second set-inclusion follows from $C(X\cup X')\subseteq X'$ (which follows from (\ref{eq:r-13})).
\end{itemize}
The above displayed inequality implies $|C(X)|>|C(X\cup X')|$. We obtain a contradiction to the law of aggregate demand. Hence, Subcase 2-1 is not possible.

\smallskip
\noindent
\emph{Subcase 2-2:}\footnote{The proof of this case is similar to that of Case 1 of step 3 in Section \ref{thm1-onlyif}.} Suppose that $X\setminus X' \nsubseteq C(X)$. Let $x\in (X\setminus X')\setminus C(X)$. 
By (\ref{eq:AMrationalization}), for any $i\in \{1, \dots, m\}$, we have $x\notin \text{\normalfont top}(X, \succ_i)=\argmax_{x\in X\cap \mathcal{X}_C}v_i(x)$. This condition implies 
\begin{align*}
u(X)=u(X-x). 
\end{align*}
By $x\notin C(X)$ and irrelevance of rejected contracts, we have $C(X-x)=C(X)$, which implies $|X\setminus C(X)|>|(X-x)\setminus C(X-x)|$. This inequality together with the above displayed equality implies that $\tilde{u}(X-x)>\tilde{u}(X)$. Therefore, condition (i) of size-restricted concavity$^+$ holds. 
\qed

\section{Proof of claims in Section \ref{sec:relation}} 
\subsection{Proof of Claim \ref{claim:submodular}} 
\label{prf:submodular-counter}
Let $\mathcal{X}=\{x,y,z\}$ and consider $u: 2^{\mathcal{X}}\rightarrow \mathbb{R}$ defined as follows: 
\begin{align*}
&u(\emptyset)=0, \: u(\{x\})=10, u(\{y\})=20, u(\{z\})=100, \\
&u(\{x,y\})=0, \: u(\{x,z\})=105, \: u(\{y,z\})=102, \: u(\{x,y,z\})=-100. 
\end{align*}

One easily verifies that $u$ is submodular. However, the choice rule $C$ rationalized by this utility function violates the substitutes condition because 
\begin{align*}
C(\{x,y,z\})\cap \{x,y\}=\{x,z\}\cap\{x,y\}=\{x\}\nsubseteq \{y\}=C(\{x,y\}). 
\end{align*}

\subsection{Proof of Claim \ref{clm:submodular-oc-independent}}
\label{prf:submodular-oc-independent}
We first show that submodularity does not imply ordinal concavity. Consider the utility function in Section \ref{prf:submodular-counter}, which satisfies submodularity. To verify that this function violates ordinal concavity, consider $\{x,z\}$, $\{y\}$, and $x\in \{x,z\}\setminus \{y\}$. We have
\begin{align*}
&u(\{x,z\})>u(\{z\}) \text{ and } u(\{y\})>u(\{x,y\}), \\
&u(\{x,z\})>u(\{y,z\}) \text{ and } u(\{y\})>u(\{x\}). 
\end{align*}
Therefore, ordinal concavity does not hold. 

To see that ordinal concavity does not imply submodularity, let $\mathcal{X}=\{x,y\}$ and define $u'$ by 
\begin{align*}
u'(\emptyset)=0, \: u'(\{x\})=1, \: u'(\{y\})=1, \: u'(\{x,y\})=3.
\end{align*}
One can verify that this function satisfies ordinal concavity. However, it violates submodularity because $u'(\{x\})+u'(\{y\})<u'(\{x,y\})+u'(\emptyset)$. 
\qed

\subsection{Proof of Claim \ref{clm:mconcave-impossibility}}
\label{prf:Mconcave-counter} 

We consider a choice rule introduced by \citet{yokoi2019}.\footnote{\citet{yokoi2019} studies a class of choice rules called {\it matroidal choice functions}. \citet{yokoi2019} proves that the choice rule given in this section is a matroidal choice function that is not rationalizable by any {\it monotonic} M$^\natural$-concave function. We strengthen this claim by showing that there exists no rationalizing M$^\natural$-concave function (including non-monotonic ones). } Let $\mathcal{X}=\{1,2,3,4,5,6\}$ and we define seven subsets of $\mathcal{X}$, denoted $Z_1$ through $Z_7$,\footnote{These seven subsets arise as cycles of an undirected graph; see Figure 6.1 of \citet{yokoi2019}.}  and $\delta(Z_k)\in \mathcal{X}$ for $k=1, \dots, 7$, as follows: 
\begin{align*}
Z_1&=\left\{1, 2, 3, 4\right\}, \delta(Z_1)=4, &\quad Z_2&=\left\{1, 3, 5, 6\right\}, \delta(Z_2)=6, \\
\quad Z_3&=\left\{2, 4, 5, 6\right\}, \delta(Z_3)=5, &Z_4&=\left\{1, 2, 6\right\}, \delta(Z_4)=6, \\
\quad Z_5&=\left\{2, 3, 5\right\}, \delta(Z_5)=5, &\quad Z_6&=\left\{3, 4, 6\right\}, \delta(Z_6)=6, \\
\quad Z_7&=\left\{1, 4, 5\right\}, \delta(Z_7)=5.
\end{align*}
Let $C$ be a choice rule given by
\begin{align*}
C(X)=X\setminus \Bigl\{\delta(Z_k)\mid Z_k\subseteq X, k\in \{1, \dots, 7\}\Bigr\} \text{ for all } X\subseteq \mathcal{X}. 
\end{align*}
This choice rule satisfies path-independence and the law of aggregate demand (see \citet{yokoi2019}). Our goal is to prove that there exists no M$^\natural$-concave utility function that rationalizes $C$. Suppose, for contradiction, that there exists a rationalizing M$^\natural$-concave function $u$. 

It is known that an M$^\natural$-concave function $u$ satisfies the following condition:\footnote{\label{footnote-Mconcave-equivalent}See, e.g., Corollary 1.3 of \citet{murotashioura2018}.} 
\begin{align*}
&\text{For any } X, X'\subseteq \mathcal{X} \text{ with } |X|=|X'| \text{ and } x\in X\setminus X', \text{there exists } x'\in X'\setminus X  \\ 
&\text{such that } u(X)+u(X')\leq u(X-x+x')+u(X'+x-x'). 
\end{align*} 
In the following, we apply the above condition to two subsets with the same size $3$ and derive $8$ equations that $u$ must satisfy. 
We abbreviate $\{x_1, x_2, \dots, x_k\}\subseteq \mathcal{X}$ to $x_1x_2\dots x_k$ for notational simplicity.
\begin{enumerate}[1.] 
\item For $X=124$ and $X'=156$ and $2\in X\setminus X'$, 
\begin{align*}
&u(124)+u(156)\leq u(126)+u(145), \text{ or } \\
&u(124)+u(156)\leq u(125)+u(146). 
\end{align*}
For $X=125$ and $X'=146$ and $2\in X\setminus X'$, 
\begin{align*}
&u(125)+u(146)\leq u(126)+u(145), \text{ or } \\
&u(125)+u(146)\leq u(124)+u(156). 
\end{align*}
By $C(1456)=146$ and $C(1256)=125$, 
\begin{align*}
u(145)<u(146) \text{ and } u(126)<u(125). 
\end{align*}
Combining the above three displayed conditions leads to 
$u(124)+u(156)=u(125)+u(146)$.\footnote{We explain why this equation holds. Applying the strict inequalities in the third displayed condition to the first displayed condition, we have $u(124)+u(156)\leq u(125)+u(146)$. Similarly, applying the strict inequalities to the second displayed condition, we have $u(125)+u(146)\leq u(124)+u(156)$. Combining these two inequalities establish the desired equation. The other 7 equations below are derived in the same manner. }   
\item For $X=134$ and $X'=156$ and $3\in X\setminus X'$, 
\begin{align*}
&u(134)+u(156)\leq u(136)+u(145), \text{ or } \\
&u(134)+u(156)\leq u(135)+u(146). 
\end{align*}
For $X=135$ and $X'=146$ and $3\in X\setminus X'$, 
\begin{align*}
&u(135)+u(146)\leq u(136)+u(145), \text{ or } \\
&u(135)+u(146)\leq u(134)+u(156). 
\end{align*}
By $C(1456)=146$ and $C(1356)=135$, 
\begin{align*}
u(145)<u(146) \text{ and } u(136)<u(135). 
\end{align*}
We obtain
$u(134)+u(156)=u(135)+u(146)$.  
\item For $X=123$ and $X'=256$ and $1\in X\setminus X'$, 
\begin{align*}
&u(123)+u(256)\leq u(126)+u(235), \text{ or } \\
&u(123)+u(256)\leq u(125)+u(236). 
\end{align*}
For $X=125$ and $X'=236$ and $1\in X\setminus X'$, 
\begin{align*}
&u(125)+u(236)\leq u(126)+u(235), \text{ or } \\
&u(125)+u(236)\leq u(123)+u(256). 
\end{align*}
By $C(2356)=236$ and $C(1256)=125$, 
\begin{align*}
u(235)<u(236) \text{ and } u(126)<u(125). 
\end{align*}
We obtain
$u(123)+u(256)=u(125)+u(236)$.  
\item 
For $X=124$ and $X'=256$ and $1\in X\setminus X'$, 
\begin{align*}
&u(124)+u(256)\leq u(126)+u(245), \text{ or } \\
&u(124)+u(256)\leq u(125)+u(246). 
\end{align*}
For $X=125$ and $X'=246$ and $1\in X\setminus X'$, 
\begin{align*}
&u(125)+u(246)\leq u(126)+u(245), \text{ or } \\
&u(125)+u(246)\leq u(124)+u(256). 
\end{align*}
By $C(2456)=246$ and $C(1256)=125$, 
\begin{align*}
u(245)<u(246) \text{ and } u(126)<u(125). 
\end{align*}
We obtain
$u(124)+u(256)=u(125)+u(246)$.  
\item 
For $X=123$ and $X'=356$ and $1\in X\setminus X'$, 
\begin{align*}
&u(123)+u(356)\leq u(136)+u(235), \text{ or } \\
&u(123)+u(356)\leq u(135)+u(236). 
\end{align*}
For $X=135$ and $X'=236$ and $1\in X\setminus X'$, 
\begin{align*}
&u(135)+u(236)\leq u(136)+u(235), \text{ or } \\
&u(135)+u(236)\leq u(123)+u(356). 
\end{align*}
By $C(2356)=236$ and $C(1356)=135$, 
\begin{align*}
u(235)<u(236) \text{ and } u(136)<u(135). 
\end{align*}
We obtain
$u(123)+u(356)=u(135)+u(236)$.  
\item 
For $X=234$ and $X'=356$ and $2\in X\setminus X'$, 
\begin{align*}
&u(234)+u(356)\leq u(236)+u(345), \text{ or } \\
&u(234)+u(356)\leq u(235)+u(346). 
\end{align*}
For $X=236$ and $X'=345$ and $2\in X\setminus X'$, 
\begin{align*}
&u(236)+u(345)\leq u(235)+u(346), \text{ or } \\
&u(236)+u(345)\leq u(234)+u(356). 
\end{align*}
By $C(3456)=345$ and $C(2356)=236$, 
\begin{align*}
u(346)<u(345) \text{ and } u(235)<u(236). 
\end{align*}
We obtain
$u(234)+u(356)=u(236)+u(345)$.  
\item 
For $X=134$ and $X'=456$ and $1\in X\setminus X'$, 
\begin{align*}
&u(134)+u(456)\leq u(146)+u(345), \text{ or } \\
&u(134)+u(456)\leq u(145)+u(346). 
\end{align*}
For $X=146$ and $X'=345$ and $1\in X\setminus X'$, 
\begin{align*}
&u(146)+u(345)\leq u(145)+u(346), \text{ or } \\
&u(146)+u(345)\leq u(134)+u(456). 
\end{align*}
By $C(3456)=345$ and $C(1456)=146$, 
\begin{align*}
u(346)<u(345) \text{ and } u(145)<u(146). 
\end{align*}
We obtain
$u(134)+u(456)=u(146)+u(345)$.  
\item 
For $X=234$ and $X'=456$ and $2\in X\setminus X'$, 
\begin{align*}
&u(234)+u(456)\leq u(246)+u(345), \text{ or } \\
&u(234)+u(456)\leq u(245)+u(346). 
\end{align*}
For $X=246$ and $X'=345$ and $2\in X\setminus X'$, 
\begin{align*}
&u(246)+u(345)\leq u(245)+u(346), \text{ or } \\
&u(246)+u(345)\leq u(234)+u(456). 
\end{align*}
By $C(3456)=345$ and $C(2456)=246$, 
\begin{align*}
u(346)<u(345) \text{ and } u(245)<u(246). 
\end{align*}
We obtain
$u(234)+u(456)=u(246)+u(345)$.  
\end{enumerate} 
Let $\mathcal{Y}$ denote the 14 subsets appearing in the above 8 equations. 
\begin{align*}
\mathcal{Y}=\{123, 124, 125, 134, 135, 146, 156, 234, 236, 246, 256, 345, 356, 456\}. 
\end{align*} 
We define $\mathcal{U}=\{\nu \in \mathbb{R}^{\mathcal{Y}} \mid A\cdot \nu=0\}$, where $A$ is the $8 \times 14$ matrix that represents the 8 equations. Specifically, $A$ is given as follows:
\begin{align*}
\begin{array}{llllllllllllll}
123 & 124 & 125 & 134 & 135 & 146 & 156 & 234 & 236 & 246 & 256 & 345 & 356 & 456 \\ \hline 
0  & 1 & -1 & 0 & 0 & -1 & 1 & 0 & 0 & 0 & 0 & 0 & 0 & 0 \\
0  & 0 & 0 & 1 & -1 & -1 & 1 & 0 & 0 & 0 & 0 & 0 & 0 & 0 \\
1  & 0 & -1 & 0 & 0 & 0 & 0 & 0 & -1 & 0 & 1 & 0 & 0 & 0 \\
0  & 1 & -1 & 0 & 0 & 0 & 0 & 0 & 0 & -1 & 1 & 0 & 0 & 0 \\
1 & 0 & 0 & 0 & -1 & 0 & 0 & 0 & -1 & 0 & 0 & 0 & 1 & 0  \\
0 & 0 & 0 & 0 & 0 & 0 & 0 & 1 & -1 & 0 & 0 & -1 & 1 & 0 \\
0 & 0 & 0 & 1 & 0 & -1 & 0 & 0 & 0 & 0 & 0 & -1 & 0 & 1 \\
0 & 0 & 0 & 0 & 0 & 0 & 0 & 1 & 0 & -1 & 0 & -1 & 0 & 1 \\
\end{array} 
\end{align*}
Letting $u|_{\mathcal{Y}}\in \mathbb{R}^{\mathcal{Y}}$ denote the vector such that $u|_{\mathcal{Y}}(Y)=u(Y)$ for all $Y\in \mathcal{Y}$, we must have $u|_{\mathcal{Y}} \in \mathcal{U}$.\footnote{\label{footnote-utility-vector}To see why this claim holds, note that we have thus far shown that any utility function $u$ satisfying M$^\natural$-concavity and rationalizing $C$ must satisfy the 8 equations. Equivalently, $u|_{\mathcal{Y}} \in \mathcal{U}$. }
One can verify the following assertion. 
\begin{fact} \label{fact1}
The matrix $A$ has full row rank. 
\end{fact}
Notice that $\mathcal{U}$ is the null space of the linear function given by the matrix $A$. Applying existing results in linear algebra, we obtain: 
\begin{proposition}[Known result in linear algebra]  \label{prop:dimension6}
$\mathcal{U}$ is a linear space and $\text{dim}(\mathcal{U})=6$.  
\end{proposition}
We define 
\begin{align*}
\mathcal{U}'=\Bigl\{\nu \in \mathbb{R}^{\mathcal{Y}} \mid &\text{ there exists } (\eta_1, \dots, \eta_6)\in \mathbb{R}^6 \text{ such that }  \\
&\: \nu(Y)=\sum_{x\in Y} \eta_x \text{ for all } Y\in \mathcal{Y}\Bigr\}.
\end{align*}  
\begin{fact} \label{fact2}
$\mathcal{U}'$ is a linear space and $\mathcal{U}'\subseteq \mathcal{U}$.
\end{fact}
\begin{proof}
It is easy to verify that $\mathcal{U}'$ is a linear space. To see that the inclusion relation holds, take $\nu\in \mathcal{U}'$. Then, there exists $(\eta_1, \dots, \eta_6)\in \mathbb{R}^6$ such that $\nu(Y)=\sum_{x\in Y} \eta_x$ for all $Y\in \mathcal{Y}$. Our goal is to show that $\nu$ satisfies the 8 equations; we only deal with the first equation because the other equations can be handled analogously. 
\begin{align*}
\nu(124)+\nu(156)&=(\eta_1+\eta_2+\eta_4)+(\eta_1+\eta_5+\eta_6) \\
&=(\eta_1+\eta_2+\eta_5)+(\eta_1+\eta_4+\eta_6)=\nu(125)+\nu(146), 
\end{align*}
as desired. 
\end{proof}
Our next goal is to prove that $\mathcal{U}'=\mathcal{U}$. To this end, we introduce two existing results. 
\begin{proposition}[Known result in linear algebra] \label{prop:space-equivalence}
For two linear spaces $\mathcal{U}$ and $\mathcal{U}'$, if $\mathcal{U}'\subseteq \mathcal{U}$ and $\text{dim}(\mathcal{U}')=\text{dim}(\mathcal{U})$, then $\mathcal{U}'=\mathcal{U}$. 
\end{proposition}
We say that two linear spaces $\mathcal{V}$ and $\mathcal{V}'$ are \textbf{isomorphic} if there exists a function $f:\mathcal{V}\rightarrow \mathcal{V}'$ such that 
\begin{itemize}
\item[1] $f$ is one-to-one. 
\item[2] $f$ is onto. 
\item[3] $f$ is linear (i.e., $f(\eta+\eta')=f(\eta)+f(\eta')$ and $f(c\cdot \eta)=c\cdot f(\eta)$). 
\end{itemize}
\begin{proposition}[Known result in linear algebra]  \label{prop:dimension} 
Two linear spaces are isomorphic if and only if they have the same dimension. 
\end{proposition}
Let $f:\mathbb{R}^6\rightarrow \mathcal{U}'$ be a function that maps $\eta\in \mathbb{R}^6$ to $\nu\in \mathcal{U}'$ given by 
\begin{align*}
\nu(Y)=\sum_{x\in Y}\eta_x \text{ for all } Y\in \mathcal{Y}.
\end{align*}
To prove that $\mathcal{U}'=\mathcal{U}$, it suffices to prove that $f$ satisfies the three conditions of isomorphism.\footnote{We explain why this claim holds. If $f$ satisfies the three conditions of isomorphism, then $\mathbb{R}^6$ and $\mathcal{U}'$ are isomorphic, which together with Proposition \ref{prop:dimension} implies $\text{dim}(\mathcal{U}')=\text{dim}(\mathbb{R}^6)=6$. This equation and Proposition \ref{prop:dimension6} implies $\text{dim}(\mathcal{U}')=\text{dim}(\mathcal{U})$. Furthermore, by Fact \ref{fact2}, $\mathcal{U}'\subseteq \mathcal{U}$. Applying Proposition \ref{prop:space-equivalence}, we obtain $\mathcal{U}'=\mathcal{U}$.} 
It is easy to see that $f$ is onto and linear. We prove that $f$ is one-to-one. 
Let $\nu \in \mathcal{U}'$ and $\eta, \eta'\in \mathbb{R}^6$ be such that $f(\eta)=f(\eta')=u$. We show that $\eta=\eta'$. By the definition of $f$, the following equation holds: 
\begin{align*}
\left(\begin{array}{cccccc}
1& 1 & 1 & 0 & 0 & 0  \\
0  & 1 & 1 & 1 & 0 & 0  \\
0  & 1 & 1 & 0 & 0 & 1  \\
1  & 0 & 0 & 1 & 0 & 1  \\
1  & 0 & 1 & 0 & 1 & 0  \\
1 & 0 & 0 & 0 & 1 & 1 
\end{array}\right)
\begin{pmatrix}
 \eta_1 \\ \eta_2 \\ \eta_3 \\ \eta_4 \\ \eta_5 \\ \eta_6 
\end{pmatrix}
=
\begin{pmatrix}
 u(123) \\ u(234) \\ u(236) \\ u(146) \\ u(135) \\ u(156) 
\end{pmatrix}. 
\end{align*}
Note that the 6 subsets appearing on the right-hand side are included in $\mathcal{Y}$. 
The same equation holds with $\eta$ replaced with $\eta'$. One can verify that the above matrix has full rank. Therefore, $\eta=\eta'$.


It follows that $\mathcal{U}=\mathcal{U}'$. Since $u|_{\mathcal{Y}}\in \mathcal{U}=\mathcal{U}'$ (recall footnote \ref{footnote-utility-vector}), 
there exists $\eta\in \mathbb{R}^6$ such that 
\begin{align}
u(Y)=\sum_{x\in Y}\eta_x \text{ for all } Y\in \mathcal{Y}.
\label{eq:utility-sum} 
\end{align}

Again, we apply M$^\natural$-concavity (in particular, the condition stated after footnote \ref{footnote-Mconcave-equivalent}) to two subsets with the same size 3. 
For $X=123$ and $X'=156$ and $2\in X\setminus X'$, 
\begin{align*}
&u(123)+u(156)\leq u(135)+u(126), \text{ or } \\
&u(123)+u(156)\leq u(136)+u(125). 
\end{align*}
Note that 126 and 136 are not included in $\mathcal{Y}$, while the other subsets are included in $\mathcal{Y}$. 
\begin{itemize}
\item If the former inequality is true, then by substituting (\ref{eq:utility-sum}) for $u(123)$, $u(156)$, and $u(135)$, we have $\eta_1+\eta_2+\eta_6\leq u(126)$. 
This inequality together with 
\begin{align*}
u(126)<u(12) \text{ (which follows from $C(126)=12$) }
\end{align*}
and
\begin{align*}
u(12)<u(125) \text{ (which follows from $C(125)=125$)} 
\end{align*}
implies $\eta_1+\eta_2+\eta_6<u(125)$. By substituting (\ref{eq:utility-sum}) for $u(125)$, we have $\eta_6<\eta_5$.   
\item If the latter inequality is true, then by substituting (\ref{eq:utility-sum}) for 123, 156, and 125, we have $\eta_1+\eta_3+\eta_6\leq u(136)$. This inequality together with 
\begin{align*}
u(136)<u(135) \text{ (which follows from $C(1356)=135$)}  
\end{align*}
implies $\eta_1+\eta_3+\eta_6<u(135)$. By substituting (\ref{eq:utility-sum}) for $u(135)$, we have $\eta_6<\eta_5$. 
\end{itemize}
Therefore, in either case, we have 
\begin{align}
\eta_6<\eta_5.
\label{eq:eta6<eta5}
\end{align} 

Applying M$^\natural$-concavity (in particular, the condition stated after footnote \ref{footnote-Mconcave-equivalent}) to $X=234$ and $X'=256$ and $3\in X\setminus X'$, we have 
\begin{align*}
&u(234)+u(256)\leq u(236)+u(245), \text{ or } \\
&u(234)+u(256)\leq u(235)+u(246). 
\end{align*}
Note that 245 and 235 are not in cluded in $\mathcal{Y}$, while the other subsets are included in $\mathcal{Y}$. 
\begin{itemize}
\item If the former inequality is true, then by substituting (\ref{eq:utility-sum}) for $u(234)$, $u(256)$, and $u(236)$, we have $\eta_2+\eta_4+\eta_5\leq u(245)$. 
This inequality together with 
\begin{align*}
u(245)<u(246) \text{ (which follows from $C(2456)=246$)}
\end{align*}
implies $\eta_2+\eta_4+\eta_5<u(246)$. By substituting (\ref{eq:utility-sum}) for $u(246)$, we have $\eta_5<\eta_6$. 
\item  If the latter inequality is true, then by substituting (\ref{eq:utility-sum}) for $u(234)$, $u(256)$, and $u(246)$, we have $\eta_2+\eta_3+\eta_5\leq u(235)$. This inequality together with 
\begin{align*}
u(235)<u(23) \text{ (which follows from $C(235)=23$)} 
\end{align*}
and 
\begin{align*}
u(23)<u(236) \text{ (which follows from $C(236)=236$)} 
\end{align*}
implies $\eta_2+\eta_3+\eta_5<u(236)$. By substituting (\ref{eq:utility-sum}) for $u(236)$, we have $\eta_5<\eta_6$.  
\end{itemize}
Therefore, in either case, we have $\eta_5<\eta_6$. We obtain a contradiction to (\ref{eq:eta6<eta5}).

\subsection{Proof of Proposition \ref{prop:pseudo}} \label{prf:pseudo} 
Let $C$ be a choice rule and $u$ be a utility function that rationalizes $C$ and satisfies pseudo M$^\natural$-concavity. Our goal is to prove that $C$ satisfies path independence and the law of aggregate demand.

\subsubsection{Proof of $C$ satisfying path independence} 
The proof is similar to that of Theorem \autoref{thm1prime} (I) in Section \ref{prf:thm1-if}. 
By Proposition \ref{prop1}, it suffices to prove that $C$ satisfies the substitutes condition and irrelevance of rejected contracts. 
One easily verifies that, if a choice rule is rationalizable by some utility function, then it satisfies irrelevance of rejected contracts. In the following we prove that $C$ satisfies the substitutes condition. 

Suppose, for contradiction, that the substitutes condition fails, i.e., there exist $X\subseteq \mathcal{X}$ and distinct $x,y\in X$ such that $x\in C(X)$ and $x\notin C(X-y)$.\footnote{We consider the negation of the condition stated in footnote \ref{footnote:sub}.} 
Considet two subsets $C(X)$, $C(X-y)$ and $x\in C(X)\setminus C(X-y)$. By pseudo M$^\natural$-concavity, there exists $x'\in \bigl(C(X-y)\setminus C(X)\bigr)\cup \{\emptyset\}$ such that 
\begin{align*}
\min\{u(C(X)), u(C(X-y))\}\leq \min\{u(C(X)-x+x'), u(C(X-y)+x-x'\}.
\end{align*}
We consider two cases. \\
\textbf{Case 1:} Suppose that the left-hand side is equal to $u(C(X))$. Then, $u(C(X))\leq u(C(X)-x+x')$. We obtain a contradiction to $C(X)$ uniquely maximizing $u(\cdot)$ among all subsets in $X$.  \\
\textbf{Case 2:} Suppose that the left-hand side is equal to $u(C(X-y))$. Then, $u(C(X-y))\leq u(C(X-y)+x-x')$. By $x\in X-y$, we obtain a contradiction to $C(X-y)$ uniquely maximizing $u(\cdot)$ among all subsets in $X-y$. 

\subsubsection{Proof of $C$ satisfying the law of aggregate demand} 
Instead of directly proving that $C$ satisfies the law of aggregate demand, we prove that $C$ satisfies the following condition: for any $X\subseteq \mathcal{X}$ and $x\in C(X)$,  
\begin{align}
C(X-x)=C(X)-x+x' \text{ for some } x'\in C(X-x)\cup\{\emptyset\}.
\label{eq:pseudo-lad-1} 
\end{align} 
This condition and irrelevance of rejected contracts of $C$ (which follows from the fact that $C$ is rationalizable by some utility function) imply that $C$ satisfies the law of aggregate demand.\footnote{To see that this implication holds, take $X, X'\subseteq \mathcal{X}$ with $X\supsetneq X'$. Let $\{x_1, \dots, x_k\}:=X\setminus X'$. We first show that $|C(X)|\geq |C(X-x_1)|$. If $x_1\notin C(X)$, then by irrelevance of rejected contracts, we have $C(X)=C(X-x_1)$, which establishes the desired inequality. If $x_1\in C(X)$, then (\ref{eq:pseudo-lad-1}) implies that the desired inequality holds. Repeating this procedure, we obtain $|C(X)|\geq |C(X-x_1)|\geq \cdots \geq |C(X-x_1-\cdots-x_k)|=|C(X')|$.}

Suppose, for contradiction, that the above condition fails, i.e., 
there exist $X\subseteq \mathcal{X}$ and $x\in C(X)$ such that
\begin{align}
C(X-x)\neq C(X)-x+x' \text{ for all } x'\in C(X-x)\cup\{\emptyset\}.
\label{eq:pseudo-lad-2} 
\end{align} 
Applying pseudo M$^\natural$-concavity to $C(X)$, $C(X-x)$ and $x\in C(X)\setminus C(X-x)$, there exists $x'\in \bigl(C(X-x)\setminus C(X)\bigr)\cup \{\emptyset\}$ such that
\begin{align*}
\min\{u(C(X)), u(C(X-x))\}\leq \min\{u(C(X)-x+x'), u(C(X-x)+x-x')\}. 
\end{align*} 
By $u(C(X))>u(C(X-x))$ (which follows from $C(X)$ uniquely maximizing $u(\cdot)$ among all subsets in $X$), the left-hand side is equal to $u(C(X-x))$. Therefore, 
\begin{align*}
u(C(X-x))\leq u(C(X)-x+x').
\end{align*}
By (\ref{eq:pseudo-lad-2}), we have $C(X-x)\neq C(X)-x+x'$. Since $x'\in C(X-x)\subseteq X-x$ (whenever $x'\neq \emptyset$), it holds that $C(X)-x+x'\subseteq X-x$. This condition and the above displayed inequality yield a contradiction to $C(X-x)$ uniquely maximizing $u(\cdot)$ among all subsets in $X-x$. 
\qed

%

%

\subsection{Proof of Claim \ref{clm:pseudo-counter}} \label{prf:pseudo-counter} 
We revisit the choice rule $C$ in Section \ref{prf:Mconcave-counter}. 
Suppose, for contradiction, that there exists a pseudo M$^\natural$-concave function that rationalizes $C$.
Applying pseudo M$^\natural$-concavity to $X=124$ and $X'=156$ and $2\in X\setminus X'$, we have 
\begin{align}
&\min\{u(124), u(156)\} \leq \min\{u(14), u(1256)\}, \text{ or } \tag*{} \\
&\min\{u(124), u(156)\} \leq \min\{u(126), u(145)\}, \text{ or } \tag*{} \\
&\min\{u(124), u(156)\} \leq \min\{u(125), u(146)\}. \label{eq:pseudo1}
\end{align}
By $u(146)>u(14)$ (which follows from $C(146)=146$) and $u(125)>u(1256)$ (which follows from $C(1256)=125$), if the first inequality holds, then the third inequality holds. By $u(125)>u(126)$ (which follows from $C(1256)=125$) and $u(146)>u(145)$ (which follows from $C(1456)=146$), if the second inequality holds, then the third inequality holds. Therefore, in any case, the third inequality (\ref{eq:pseudo1}) holds. 

Applying pseudo M$^\natural$-concavity to $X=125$ and $X'=146$ and $2\in X\setminus X'$, we have  
\begin{align}
&\min\{u(125), u(146)\} \leq \min\{u(15), u(1246)\}, \text{ or } \tag*{} \\
&\min\{u(125), u(146)\} \leq \min\{u(126), u(145)\}, \text{ or } \tag*{} \\
&\min\{u(125), u(146)\} \leq \min\{u(124), u(156)\}.  \label{eq:pseudo2}
\end{align}
By $u(156)>u(15)$ (which follows from $C(156)=156$) and $u(124)>u(1246)$ (which follows from $C(1246)=124$), if the first inequality holds, then the third inequality holds. By $u(125)>u(126)$ (which follows from $C(1256)=125$) and $u(146)>u(145)$ (which follows from $C(1456)=146$), the second inequality never holds. Therefore, the third inequality (\ref{eq:pseudo2}) holds. 

Combining (\ref{eq:pseudo1}) and (\ref{eq:pseudo2}), we have 
\begin{align*}
\min\{u(124), u(156)\}=\min\{u(125), u(146)\}.
\end{align*}  
By $u(124)>u(156)$ (which follows from $C(12456)=124$), the left-hand side is equal to $u(156)$. 
If $u(156)=u(125)$, then we obtain a contradiction to $u(125)>u(156)$ (which follows from $C(1256)=125$). If $u(156)=u(146)$, then we obtain a contradiction to $u(146)>u(156)$ (which follows from $C(1456)=146$). 
\qed

\newpage

\setcounter{page}{1} 

\begin{center}
\textbf{Online Appendix}

\textbf{Alternative Proof of Theorem \ref{thm1}}
\end{center}
In this section, we provide an alternative proof of the only-if direction of Theorem \ref{thm1}.
Let $\cf$ be a choice rule that satisfies path independence. Then it also satisfies irrelevance of rejected contracts and the substitutes condition (Proposition \ref{prop1}).

We proceed in three steps.
In Section \ref{ftilde-derivation}, we construct a utility function $\tilde u$.
In Section \ref{ftilde-represent}, we prove that $\tilde u$ rationalizes $C$. In Section \ref{ftilde-ordinal}, we prove that $\tilde u$ satisfies ordinal concavity.

\subsection{Construction of a utility function} \label{ftilde-derivation}
Let $n=|\mathcal{X}|$. For any $X\subseteq \mathcal{X}$, we define $E_X^k$ for $k=1, \dots, n$, inductively as follows:
\begin{align*}
E_X^1&=\mathcal{X}, \\
E_X^k&=E_X^{k-1}\setminus (\cf(E_X^{k-1})\setminus X) \: \text{ for } k=2, \dots, n.
\end{align*}
We define $\alpha_k$ for $k=1, \dots, n$, inductively (from $n$ to $1$) as follows:
\begin{align*}
\alpha_n&=1, \\
\alpha_k&=\max_{X\subseteq \mathcal{X}} \sum_{j=k+1}^{n} \alpha_{j} \cdot |X\cap \cf(E_X^{j})|+1
\: \text{ for } k=n-1, n-2, \dots, 1.
\end{align*}
We define $u: 2^{\mathcal{X}}\rightarrow \mathbb{R}$ by
\begin{align}
u(X)=\sum_{k=1}^{n} \alpha_k \cdot |X\cap \cf(E_X^k)| \text{ for every } X\subseteq \mathcal{X}.
\label{eq:represent-0}
\end{align}
For any $X\subseteq \mathcal{X}$ and $k=1, \dots, n$, we define $\delta_X^k$ by
\begin{align*}
\delta_X^k=\begin{cases}  \varepsilon & \text{ if } \cf(E_X^k)\subseteq X \text{ and } \cf(E_X^j)\nsubseteq X \text{ for every } j \text{ with } j<k, \\
                                     0 & \text{ otherwise, }
                 \end{cases}
\end{align*}
where $\varepsilon$ is a sufficiently small number with $0<\varepsilon<1/n$.
We define $\tilde u: 2^{\mathcal{X}}\rightarrow \mathbb{R}$ by
\begin{align}
\tilde u(X)=
u(X)-\sum_{k=1}^n \delta_X^k\cdot |X\setminus \cf(E_X^k)| \text{ for every } X\subseteq \mathcal{X}.
\label{eq:represent-1}
\end{align}
The following inequalities hold:
\begin{align}
u(X)\geq \tilde u(X) \text{ and } \tilde u(X)>u(X)-1 \text{ for every } X\subseteq \mathcal{X},
\label{tilde-strict}
\end{align}
where the strict inequality follows from $\sum_{k=1}^n \delta_X^k\cdot |X\setminus \cf(E_X^k)|\leq \varepsilon \cdot n<1$.

\begin{claim} \label{claim:representation}
Let $X, X'\subseteq \mathcal{X}$ with $X\subseteq X'$. Then,
\begin{align*}
E_X^j\subseteq E_{X'}^j \text{ for every } j=1, \dots, n.
\end{align*}
\end{claim}
\begin{proof}
The proof is by mathematical induction. The claim trivially holds for $j=1$ because $E_X^1=E_{X'}^1=\mathcal{X}$. Suppose that it holds for $j-1$. We show the claim for $j$.

By the definition of $E$, our goal is to prove that
\begin{align}
\Bigl(E_X^{j-1}\setminus (\cf(E_X^{j-1})\setminus X)\Bigr)  \subseteq \Bigl(E_{X'}^{j-1}\setminus (\cf(E_{X'}^{j-1})\setminus X')\Bigr).
\label{11-22-5}
\end{align}
Let $x\in E_X^{j-1}\setminus (\cf(E_X^{j-1})\setminus X)$. By $x\in E_X^{j-1}$ and the induction hypothesis,
\begin{align}
x\in E_{X'}^{j-1}.
\label{11-22-6}
\end{align}
By $x\notin \cf(E_X^{j-1})\setminus X$, we have (i) $x\notin \cf(E_X^{j-1})$ or (ii) $x\in \cf(E_X^{j-1})\cap X$. If (i) holds, then by $x\in E_X^{j-1}$, the induction hypothesis, and the substitutes condition,\footnote{We use the following equivalent condition to the substitutes condition: for any $X, X' \subseteq \calx$ with $X\subseteq X'$, it holds that $X\backslash C(X)\subseteq X'\backslash C(X')$.}
we have $x\notin \cf(E_{X'}^{j-1})$, which implies $x\notin \cf(E_{X'}^{j-1})\setminus X'$.
Together with (\ref{11-22-6}), it implies that $x$ is included in the right-hand side of (\ref{11-22-5}). If (ii) holds, 
then $x\in X$, which implies $x\in X'$, and so $x\notin \cf(E_{X'}^{j-1})\setminus X'$. Together with (\ref{11-22-6}),
it implies that $x$ is included in the right-hand side of (\ref{11-22-5}).
%
\end{proof}

\subsection{Proof of $\tilde u$ rationalizing $C$}
\label{ftilde-represent}

Fix an arbitrary $\bar{X}\subseteq \mathcal{X}$ with $\bar{X}\neq \emptyset$ and
let $X^*=\cf(\bar{X})$.
Our goal is to prove that $X^*$ uniquely maximizes $\tilde u$ among all subsets of $\bar{X}$.
The proof works in a number of steps. At each step $k$, where $1\leq k \leq n$, we provide two statements labeled as $(a|k)$ and $(b|k)$ below.

In step 1, we show that either Claim $(a|1)$ or Claim $(b|1)$ holds. The proof is completed if $(a|1)$ holds. If $(b|1)$ holds, then we go to step 2.
In step 2, we show that either Claim $(a|2)$ or Claim $(b|2)$ holds. Again, if $(a|2)$ holds then the proof is completed, and otherwise we go to
step 3. We continue this process until Claim $(a|k)$ holds for some $k\in \{1, \dots, n\}$. 

We define $\Psi^k$ for $k=0, \dots, n$ inductively as follows:
\begin{align*}
\Psi^0&=\{X\subseteq \mathcal{X} \mid X\subseteq \bar{X}\}, \\
\Psi^k&=\bigl\{X\in \Psi^{k-1} \mid |X\cap \cf(E_{\bar{X}}^k)|\geq |X'\cap \cf(E_{\bar{X}}^k)| \: \text{ for every } X'\in \Psi^{k-1}\bigr\} \\
&\hspace{83.5mm} \text{ for every } k=1, \dots, n.
\end{align*}
Note that $\Psi^0\supseteq \Psi^1 \supseteq \cdots \supseteq \Psi^n$.

\medskip
\paragraph{\textbf{Step \boldmath{$k$ ($1\leq k\leq n$)}.}}
Suppose that one of the following two conditions holds:
\begin{itemize}
\item $k=1$, or
\item $k\geq 2$ and $(b|j)$ holds in every step $j=1, \dots, k-1$.
\end{itemize}
Then, one of the following two claims holds:
\begin{description}
\item[$(a|k)$]  $X^*$ uniquely maximizes $\tilde u$ among all elements in $\Psi^{0}$; or
\item[$(b|k)$] $\delta_X^k=0$ for every $X\in \Psi^{k-1}$, and $\Psi^k$ satisfies the following four conditions:
    \begin{enumerate}
      \item[(i)] $X^*\in \Psi^k$,
      \item[(ii)] $X\cap \cf(E_{X}^k)=\bar{X} \cap \cf(E_{\bar{X}}^k)$ for every $X\in \Psi^k$,
      \item[(iii)] $E_X^{k+1}=E_{\bar{X}}^{k+1}\subsetneq E_{\bar{X}}^k$ for every $X\in \Psi^k$, and
      \item[(iv)] $u(X)>u(X')$ for every $X\in \Psi^k$ and $X'\in \Psi^{k-1} \setminus \Psi^k$.
    \end{enumerate}
\end{description}
Moreover, if $k=n$, then $(a|n)$ holds.
\begin{proof}[Proof of the statement for step $k$]\renewcommand{\qedsymbol}{$\blacksquare$}
By the definition of $E$, 
$E_{\bar{X}}^k \supseteq \bar{X}$. Together with the substitutes condition, it implies that
\begin{align}
X^* = \cf(\bar{X}) \supseteq \bar{X} \cap \cf(E_{\bar{X}}^k).
\label{2-14-1-k}
\end{align}
The following equality holds:
\begin{align}
&\Psi^k=\bigl\{X\in \Psi^{k-1} \mid X\supseteq \bar{X}\cap \cf(E_{\bar{X}}^k)\bigr\}.
\label{11-19-1-k}
\end{align}
To see that (\ref{11-19-1-k}) holds, let $X \in \Psi^{k-1}$ with $X\supseteq \bar{X}\cap \cf(E_{\bar{X}}^k)$. Then, $X \cap \cf(E_{\bar{X}}^k) \supseteq \bar{X}\cap \cf(E_{\bar{X}}^k)$, which implies $|X \cap \cf(E_{\bar{X}}^k)|\geq |\bar{X}\cap \cf(E_{\bar{X}}^k)|$. Since $|\bar{X}\cap \cf(E_{\bar{X}}^k)| \geq |X'\cap \cf(E_{\bar{X}}^k)|$ for every $X'\in \Psi^{k-1}\subseteq \Psi^0$, we obtain $X\in \Psi^k$. Conversely, let $X\in \Psi^{k-1}$ with $X\nsupseteq \bar{X}\cap \cf(E_{\bar{X}}^k)$. Then,
\begin{align*}
|X^*\cap C(E_{\bar{X}}^k)|=|X^*\cap \bigl(\bar{X}\cap C(E_{\bar{X}}^k)\bigr)|>|X\cap \bigl(\bar{X}\cap C(E_{\bar{X}}^k)\bigr)|=|X\cap C(E_{\bar{X}}^k)|,
\end{align*}
where the first equality follows from $X^*\subseteq \bar{X}$, the strict inequality follows from (\ref{2-14-1-k}) and $X\nsupseteq \bar{X}\cap \cf(E_{\bar{X}}^k)$, and the last equality follows from $X\subseteq \bar{X}$. By the above displayed inequality and $X^*\in \Psi^{k-1}$ (which follows from $(b|k-1)$ (i)),\footnote{If $k=1$, then $X^*\in \Psi^0$ follows from $X^*=C(\bar{X})\subseteq \bar{X}$.} we get $X\notin \Psi^k$. It follows that (\ref{11-19-1-k}) holds.

By (\ref{2-14-1-k}), (\ref{11-19-1-k}) and $X^*\in \Psi^{k-1}$,
\begin{align}
&X^*\in \Psi^k.
\label{10-27-1-k}
\end{align}

For any $X\in \Psi^k$ and $X'\in \Psi^{k-1}\setminus \Psi^k$,\footnote{If $k=n$, then the summation $\sum_{j=k+1}^n \alpha_j \cdot |X\cap C(E_X^j)|$ is defined to be $0$.}
\begin{align*}
&u(X)-u(X') \tag*{} \\
&=\Bigl\{\sum_{j=1}^k \alpha_j\cdot |X\cap \cf(E_{X}^j)|+\sum_{j=k+1}^{n}\alpha_j \cdot |X\cap \cf(E_{X}^j)|\Bigr\} \tag*{} \\
&-\Bigl\{\sum_{j=1}^k\alpha_j \cdot |X'\cap \cf(E_{X'}^j)|+\sum_{j=k+1}^{n}\alpha_j \cdot |X'\cap \cf(E_{X'}^j)| \Bigr\}
\tag*{} \\
&\geq \sum_{j=1}^k \alpha_j \cdot |X\cap \cf(E_{X}^j)|-\Bigl\{\sum_{j=1}^k\alpha_j\cdot |X' \cap \cf(E_{X'}^j)|+\sum_{j=k+1}^{n} \alpha_j \cdot |X'\cap \cf(E_{X'}^j)|\Bigr\} \tag*{} \\
&=\alpha_k \cdot |X\cap \cf(E_{X}^k)|-\alpha_k\cdot |X' \cap \cf(E_{X'}^k)|-\sum_{j=k+1}^{n} \alpha_j \cdot |X'\cap \cf(E_{X'}^j)| \tag*{} \\
&=\alpha_k \cdot |X\cap \cf(E_{\bar{X}}^k)|-\alpha_k\cdot |X' \cap \cf(E_{\bar{X}}^k)|-\sum_{j=k+1}^{n} \alpha_j \cdot |X'\cap \cf(E_{X'}^j)| \tag*{} \\
&\geq \alpha_k-\sum_{j=k+1}^{n} \alpha_j \cdot |X'\cap \cf(E_{X'}^j)| \tag*{} \\
&\geq 1,
\end{align*}
where 
\begin{itemize}
\item the first inequality follows from $\sum_{j=k+1}^{n}\alpha_j \cdot |X \cap \cf(E_{X}^j)|\geq 0$,
\item the second equality follows from $(b|j)$ (ii) for every $j$ with $1\leq j \leq k-1$,\footnote{If $k=1$, then the second equality trivially holds.}
\item the third equality follows from $E_X^1=E_{X'}^1=E_{\bar{X}}^1=\mathcal{X}$ (if $k=1$), and $X,X'\in \Psi^{k-1}$ and $(b|k-1)$ (iii) (if $k\geq 2$),
\item the second inequality follows from $X \in \Psi^k$ and $X' \notin \Psi^k$, and
\item the last inequality follows from the definition of $\alpha_k$.
\end{itemize}
Hence,
\begin{align}
u(X)-u(X')\geq 1. \label{2-14-2-k}
\end{align}
We consider two cases. \\
\textbf{Case 1:} Suppose $\bar{X} \supseteq \cf(E_{\bar{X}}^k)$. We prove that $(a|k)$ holds.

By $\cf(E_{\bar{X}}^k)\subseteq \bar{X}\subseteq E_{\bar{X}}^k$ and irrelevance of rejected contracts,
\begin{align}
X^*=\cf(\bar{X}) = \cf(E_{\bar{X}}^k).
\label{10-27-3-k}
\end{align}
Fix $X\in \Psi^k$ with $X\neq X^*$. The above equality and 
(\ref{11-19-1-k}) imply
\begin{align*}
X\supseteq \bar{X}\cap C(E_{\bar{X}}^k) = \bar{X}\cap X^*=X^*=\cf(E_{\bar{X}}^k).
\end{align*}
Together with (\ref{10-27-3-k}) and $X\neq X^*$, it implies
\begin{align}
X\supsetneq \cf(E_{\bar{X}}^k).
\label{10-24-1-k}
\end{align}
By (\ref{10-27-1-k}), $X\in \Psi^k$, and $(b|k-1)$ (iii),\footnote{If $k=1$, then (\ref{11-6-1-k}) follows from $E_{X^*}^1=E_{X}^1=E_{\bar{X}}^1=\mathcal{X}$.} we have
\begin{align}
E_{X^*}^k=E_{X}^k=E_{\bar{X}}^k.
\label{11-6-1-k}
\end{align}
Together with \eqref{10-27-3-k} and \eqref{10-24-1-k}, it yields
\begin{align}
X^*=\cf(E_{X^*}^k), \: X\supsetneq \cf(E_{X}^k),
\label{10-27-4-k}
\end{align}
which implies $\cf(E_{X^*}^k)\setminus X^*=\emptyset$ and $\cf(E_{X}^k)\setminus X=\emptyset$. By the definition of $E$, 
\begin{align*}
E_{X^*}^{k}=E_{X^*}^{k+1}=\dots=E_{X^*}^n \text{ and } E_X^k=E_X^{k+1}=\dots=E_X^n.
\end{align*}
Together with (\ref{11-6-1-k}) and (\ref{10-27-4-k}), it implies
\begin{align}
X^*\cap \cf(E_{X^*}^j)=X \cap \cf(E_{X}^j) \: \text{ for every } j=k, \dots, n.
\label{11-6-2-k}
\end{align}
By $(b|j)$ (ii) for every $j$ with $1\leq j \leq k-1$,\footnote{If $k=1$, then we do not need (\ref{11-6-3-k}) in order to establish (\ref{10-27-5}).}
\begin{align}
X^*\cap \cf(E_{X^*}^j)=X \cap \cf(E_{X}^j) \: \text{ for every } j \text{ with } 1\leq j \leq k-1.
\label{11-6-3-k}
\end{align}
By (\ref{11-6-2-k}) and (\ref{11-6-3-k}),
\begin{align}
u(X^*)=u(X).
\label{10-27-5}
\end{align}
By the first claim of $(b|j)$ for every $j$ with $1\leq j \leq k-1$, \eqref{10-27-4-k}, and the definition of $\delta$, 
\begin{align}
\delta_{X^*}^k=\delta_{X}^k=\varepsilon \text{ and } \delta_{X^*}^j=\delta_{X}^j=0 \: \text{ for every } j\in \{1, \dots, n\} \text{ with } j\neq k.
\label{10-27-6-k}
\end{align}
We obtain
\begin{align}
\tilde{u}(X^*)=u(X^*)-\delta_{X^*}^k\cdot |X^*\setminus \cf(E_{X^*}^k)|&=u(X^*) \label{tilde-equality-k} \\
&>u(X)-\delta_{X}^k\cdot |X\setminus \cf(E_{X}^k)| \tag*{} \\
&=\tilde{u}(X), \label{11-12-1-k}
\end{align}
where the first and last equalities follow from (\ref{10-27-6-k}), the second equality follows from (\ref{10-27-4-k}), and the strict inequality follows from (\ref{10-27-4-k}) and (\ref{10-27-5}).

Furthermore, for any $j=0, \dots, k-1$ and any $X^j\in \Psi^j\backslash \Psi^{j+1}$,
\begin{align}
u(X^*)>u(X^{k-1})>\dots>u(X^{0}),
\label{11-12-2}
\end{align}
where the first inequality follows from \eqref{10-27-1-k} and \eqref{2-14-2-k} and the other inequalities follow from $(b|j)$ (iv) for every $j$ with $1\leq j \leq k-1$.
Hence, for any $X' \in \Psi^{0}\setminus \Psi^k$, we have $\tilde u(X^*)=u(X^*)>u(X')\geq \tilde u(X')$, where the equality follows from (\ref{tilde-equality-k}), the strict inequality follows from (\ref{11-12-2}),
and the weak inequality follows from the former inequality of (\ref{tilde-strict}).
Together with (\ref{11-12-1-k}) for every $X\in \Psi^k$ with $X\neq X^*$, it yields $(a|k)$.
\\
\textbf{Case 2:} Suppose $\bar{X} \nsupseteq \cf(E_{\bar{X}}^k)$. We prove that $(b|k)$ holds. For any $X\in \Psi^{k-1}$, by the assumption of Case 2 and $X\subseteq \bar{X}$ (which follows from $X\in \Psi^{k-1}\subseteq \Psi^0$), we have $\cf(E_X^k)=\cf(E_{\bar{X}}^k)\nsubseteq X$ (where the equality follows from $(b|k-1)$ (iii)),\footnote{If $k=1$, then the equality follows from $E_X^1=E_{\bar{X}}^1=\mathcal{X}$. The same comment applies to the equation $E_X^k=E_{\bar{X}}^k$ in the remaining part.} which implies $\delta_X^k=0$. Thus,
the first claim of $(b|k)$ holds.

By (\ref{10-27-1-k}), we obtain $(b|k)$ (i).

For any $X\in \Psi^k$, 
by (\ref{11-19-1-k}), we have $X \supseteq \bar{X} \cap \cf(E_{\bar{X}}^k)$. Together with $X\subseteq \bar{X}$ (which follows from $X\in \Psi^k\subseteq \Psi^0$),
 it yields
%
\begin{align}
\bar{X} \cap \cf(E_{\bar{X}}^k)=X\cap \cf(E_{\bar{X}}^k).
\label{10-26-1-k}
\end{align}
Together with $E_X^k=E_{\bar{X}}^k$ (which follows from $(b|k-1)$ (iii)),
it implies $(b|k)$ (ii).

For any $X\in \Psi^k$, we have
\begin{align*}
E_X^{k+1}&=E_{X}^k\setminus (\cf(E_{X}^k)\setminus X) \\
&=E_{\bar{X}}^k\setminus (\cf(E_{\bar{X}}^k)\setminus X) \\
&=E_{\bar{X}}^k\setminus \Bigl(\cf(E_{\bar{X}}^k)\setminus \bigl(X\cap \cf(E_{\bar{X}}^k)\bigr)\Bigr) \\
&=E_{\bar{X}}^k\setminus \Bigl(\cf(E_{\bar{X}}^k)\setminus \bigl(\bar{X}\cap \cf(E_{\bar{X}}^k)\bigr)\Bigr) \\
&=E_{\bar{X}}^k\setminus (\cf(E_{\bar{X}}^k)\setminus \bar{X}) \\
&=E_{\bar{X}}^{k+1}.
\end{align*}
where the second equality follows from $E_X^k=E_{\bar{X}}^k$ (which follows from $(b|k-1)$ (iii)) and the fourth equality follows from (\ref{10-26-1-k}).
Together with $\cf(E_{\bar{X}}^k)\setminus \bar{X}\neq \emptyset$ (which follows from the assumption of Case 2), it yields $(b|k)$ (iii).
Finally, $(b|k)$ (iv) follows from \eqref{2-14-2-k}.
We conclude that $(b|k)$ holds.

Finally, we prove the claim that $(a|k)$ holds for $k=n$. By $(b|j)$ (iii) for $j$ with $1 \leq j \leq n-1$, we get $|E^n_{\bar{X}}|\leq 1$. Together with $\bar{X}\neq \emptyset$ and $\bar{X}\subseteq E^n_{\bar{X}}$ (which follows from the definition of $E$), we get $E^n_{\bar{X}}=\bar{X}$. Hence, $C(E^n_{\bar{X}})\subseteq \bar{X}$. As proven in Case 1, $(a|n)$ holds.
\end{proof}

By the statement for step $k$ ($1\leq k\leq n$), there exists $j\in \{1, \dots, n\}$ such that $(a|j)$ holds. Therefore, we obtain the desired claim.  \qed

\subsection{Proof of ordinal concavity of $\tilde u$} \label{ftilde-ordinal}
As in the proof of Theorem \autoref{thm1prime} (I) in the main text, we show that $\tilde u$ satisfies ordinal concavity$^{++}$ (see step 3 in Section \ref{thm1-onlyif}). Since ordinal concavity$^{++}$ is stronger than ordinal concavity (see footnote \ref{footnote:oc++} in the main text), the desired claim follows. 


\begin{defa}[Same as Definition \ref{def:oc+} in the main text] 
A utility function $u$ satisfies \textbf{ordinal concavity$^{++}$} if, for any $X, X'\subseteq \mathcal{X}$ and $x\in X\setminus X'$, one of the following two conditions holds:
\begin{enumerate}[(i)]
\item there exists $x'\in (X'\setminus X)\cup \{\emptyset\}$ such that $u(X)<u(X-x+x')$, or
\item $u(X')<u(X'+x)$.
\end{enumerate}
\end{defa}

Let $X, X'\subseteq \mathcal{X}$ and $x\in X\setminus X'$. We show that 
$\tilde u$ defined by (\ref{eq:represent-1}) satisfies (i) or (ii) in the above definition.

Let $\bar{X}=X\cup X'$ and $X^*=C(\bar{X})$. By following the same line of the proof in Section \ref{ftilde-represent},
there exists $\ell\in \{0, \dots, n-1\}$ such that
\begin{itemize}
\item For every $j$ with $1\leq j\leq \ell$, Case 2 holds in step $j$ and we obtain $(b|j)$, and
\item Case 1 holds in step $\ell+1$ and we obtain $(a|\ell+1)$.
\end{itemize}
As shown in the proof, 
there exists a sequence of collections of subsets of $\bar{X}$, $(\Psi^0, \dots, \Psi^\ell)$, such that $\Psi^j$ satisfies the conditions in $(b|j)$ for every $j$ with $1\leq j\leq \ell$.
We define $k(X), k(X')\in \{0, \dots, \ell\}$ by
\begin{align*}
k(X)&=\max\bigl\{j\in \{0, \dots, \ell\} \mid X\in \Psi^j\bigr\}, \\
k(X')&=\max\bigl\{j\in \{0, \dots, \ell\} \mid X'\in \Psi^j\bigr\}.
\end{align*}
Let $\bar{k}=\min\{k(X), k(X')\}$.
By (\ref{11-19-1-k}) and $X,X'\in \Psi^j$ for every $j$ with $1\leq j\leq \bar{k}$,
\begin{align*}
X\supseteq \bar{X}\cap \cf(E^j_{\bar{X}}) \text{ and } X'\supseteq \bar{X}\cap \cf(E^j_{\bar{X}}) \text{ for every } j \text{ with } 1\leq j \leq \bar{k}.
\end{align*}
Together with $x\in X\setminus X'$, it implies
\begin{align*}
&X-x+x'\supseteq \bar{X}\cap \cf(E^j_{\bar{X}}) \text{ for every } x'\in (X'\setminus X)\cup\{\emptyset\} \text{ and } j \text{ with } 1\leq j \leq \bar{k}, \text{ and } \\
&X'+x\supseteq \bar{X}\cap \cf(E^j_{\bar{X}}) \text{ for every } j \text{ with } 1\leq j \leq \bar{k}.
\end{align*}
These conditions and (\ref{11-19-1-k}) imply
\begin{align}
&X-x+x' \in \Psi^{j} \text{ for every } x'\in (X'\setminus X)\cup\{\emptyset\} \text{ and } j \text{ with } 1\leq j \leq \bar{k}, \text{ and }
\label{11-19-3*} \\
&X'+x \in \Psi^{j} \text{ for every }  j \text{ with } 1\leq j \leq \bar{k}. \label{11-23-1}
\end{align}
We consider two cases. \\
\textbf{Case 1:}
Suppose $k(X')>k(X)$. By the definition of $k(X)$, we have $X\in \Psi^{k(X)}$ and $X\notin \Psi^{k(X)+1}$. Together with $X'\in \Psi^{k(X)+1}$ and (\ref{11-19-1-k}),
it yields
\begin{align*}
X'\supseteq \bar{X}\cap \cf(E^{k(X)+1}_{\bar{X}}) \text{ and } X\nsupseteq \bar{X}\cap \cf(E^{k(X)+1}_{\bar{X}}).
\end{align*}
Since $\bar{X}=X\cup X'$, the above conditions have two implications. First, the former set-inclusion and $x\in X\setminus X'$ imply
\begin{align}
x\notin \cf(E^{k(X)+1}_{\bar{X}}).
\label{11-19-2}
\end{align}
Second, there exists $x'\in X'\setminus X$ such that
\begin{align}
x'\in \cf(E^{k(X)+1}_{\bar{X}}).
\label{11-19-2*}
\end{align}
By \eqref{11-19-3*} and $x'\in X'\setminus X$,
\begin{align}
X-x+x' \in \Psi^{j} \text{ for every } j \text{ with } 1\leq j \leq k(X).
\label{11-19-3}
\end{align}
We obtain
\begin{align}
&u(X-x+x')-u(X) \tag*{} \\
&=\Bigl\{\sum_{j=1}^{k(X)+1} \alpha_j\cdot |(X-x+x')\cap \cf(E_{X-x+x'}^j)|+\sum_{j=k(X)+2}^{n}\alpha_j \cdot |(X-x+x')\cap \cf(E_{X-x+x'}^j)|\Bigr\} \tag*{} \\
&-\Bigl\{\sum_{j=1}^{k(X)+1}\alpha_j \cdot |X\cap \cf(E_{X}^j)|+\sum_{j=k(X)+2}^{n}\alpha_j \cdot |X\cap \cf(E_{X}^j)| \Bigr\}
\tag*{} \\
&\geq \sum_{j=1}^{k(X)+1} \alpha_j \cdot |(X-x+x')\cap \cf(E_{X-x+x'}^j)| \tag*{} \\
&-\Bigl\{\sum_{j=1}^{k(X)+1}\alpha_j\cdot |X \cap \cf(E_{X}^j)|+\sum_{j=k(X)+2}^{n} \alpha_j \cdot |X\cap \cf(E_{X}^j)|\Bigr\} \tag*{} \\
&=\alpha_{k(X)+1} \cdot |(X-x+x')\cap \cf(E_{X-x+x'}^{k(X)+1})| \tag*{} \\
&-\alpha_{k(X)+1}\cdot |X \cap \cf(E_{X}^{k(X)+1})|-\sum_{j=k(X)+2}^{n} \alpha_j \cdot |X\cap \cf(E_{X}^j)| \tag*{} \\
&=\alpha_{k(X)+1} \cdot |(X-x+x')\cap \cf(E_{\bar{X}}^{k(X)+1})| \tag*{} \\
&-\alpha_{k(X)+1}\cdot |X \cap \cf(E_{\bar{X}}^{k(X)+1})|-\sum_{j=k(X)+2}^{n} \alpha_j \cdot |X\cap \cf(E_{X}^j)| \tag*{} \\
&=\alpha_{k(X)+1}-\sum_{j=k(X)+2}^{n} \alpha_j \cdot |X\cap \cf(E_{X}^j)| \tag*{} \\
&\geq 1,  \tag*{} 
\end{align}
where
\begin{itemize}
\item the first inequality follows from $\sum_{j=k(X)+2}^{n}\alpha_j \cdot |(X-x+x')\cap \cf(E_{X-x+x'}^j)|\geq 0$,
\item the second equality follows from $X-x+x'\in \Psi^j$ (which is implied by (\ref{11-19-3})), $X\in \Psi^j$, and $(b|j)$ (ii) for every $j$ with $1\leq j \leq k(X)$,
\item the third equality follows from $X-x+x'\in \Psi^{k(X)}$ (which is implied by (\ref{11-19-3})), $X\in \Psi^{k(X)}$, and $(b|k(X))$ (iii),\footnote{If $k(X)=0$, then the third equality follows from $E_{X-x+x'}^1=E_X^1=E_{\bar{X}}^1=\mathcal{X}$.}
\item the fourth equality follows from (\ref{11-19-2}) and (\ref{11-19-2*}), and
\item the last inequality follows from the definition of $\alpha_{k(X)+1}$.
\end{itemize}
It follows that $u(X-x+x')\geq u(X)+1$, which together with (\ref{tilde-strict}) implies $\tilde u(X-x+x')>\tilde u(X)$. Hence, condition (i) of ordinal concavity$^{++}$ holds.
\\
\textbf{Case 2:} Suppose $k(X')\leq k(X)(\leq n-1)$. We consider two subcases. \\
\textbf{Subcase 2-1:} Suppose that $x\notin \cf(E_{\bar{X}}^k)$ for every $k\in \{k(X')+1, \dots, n\}$.

By (\ref{11-19-1-k}) and the definition of $k(X)$, we have $X\supseteq \bar{X}\cap \cf(E^j_{\bar{X}})$ for every $j$ with $k(X')+1\leq j \leq k(X)$.
Together with the assumption of Subcase 2-1, it yields
\begin{align*}
X-x+x'&\supseteq \bar{X}\cap \cf(E^j_{\bar{X}}) \\
&\text{ for every } x'\in (X'\setminus X)\cup\{\emptyset\} \text{ and } j \text{ with } k(X')+1\leq j \leq k(X).
\end{align*}
This condition and (\ref{11-19-1-k}) imply
\begin{align*}
X-x+x' \in \Psi^{j} \text{ for every } x'\in (X'\setminus X)\cup\{\emptyset\} \text{ and } j \text{ with } k(X')+1\leq j \leq k(X).
\end{align*}
These conditions, together with (\ref{11-19-3*}), yield
\begin{align}
X-x+x' \in \Psi^{j} \text{ for every } x'\in (X'\setminus X)\cup\{\emptyset\} \text{ and } j \text{ with } 1\leq j\leq k(X).
\label{11-22-1}
\end{align}
We consider two further subcases. \\
\textbf{Subcase 2-1-1:} Suppose $k(X)=\ell$.
As we noted in the beginning of Section \ref{ftilde-ordinal} (after the definition of ordinal concavity$^{++}$), Case 1 holds in the proof of step $\ell+1$ in Section \ref{ftilde-represent}, which
implies $\bar{X}\supseteq \cf(E_{\bar{X}}^{\ell+1})$ (see the first sentence of Case 1 after (\ref{2-14-2-k})).
We consider two further subcases.
\\
\textbf{Subcase 2-1-1-1:} Suppose $X\supseteq \cf(E_{\bar{X}}^{\ell+1})$.
By $X-x\in \Psi^\ell$ (which follows from (\ref{11-22-1})), $X\in \Psi^\ell$, and $(b|\ell)$ (iii),\footnote{If $\ell=0$, then (\ref{11-20-0}) follows from $E^1_{X-x}=E^1_X=E^1_{\bar{X}}=\mathcal{X}$.}
we have
\begin{align}
E_{X-x}^{\ell+1}=E_{X}^{\ell+1}=E_{\bar{X}}^{\ell+1}.
\label{11-20-0}
\end{align}
Together with the assumptions of Subcases 2-1 and 2-1-1-1, these equations imply
\begin{align}
X-x\supseteq \cf(E_{X-x}^{\ell+1}) \text{ and } X\supseteq \cf(E_{X}^{\ell+1}).
\label{12-2-2}
\end{align}
If $\ell+1<n$, then by the definition of $E$ 
and (\ref{12-2-2}), we get
\begin{align*}
E_{X-x}^{\ell+1}=E_{X-x}^{\ell+2} \text{ and } E_{X}^{\ell+1}=E_{X}^{\ell+2},
\end{align*}
which together with (\ref{12-2-2}) imply $X-x\supseteq \cf(E_{X-x}^{\ell+2})$ and $X\supseteq \cf(E_{X}^{\ell+2})$. Repeating this procedure,
\begin{align}
&E_{X-x}^{j}=E_{X}^{j} \text{ for every } j=\ell+1, \dots, n, \text{ and } \label{11-20-1} \\
&X-x\supseteq \cf(E_{X-x}^{j}) \text{ and } X\supseteq \cf(E_{X}^{j}) \text{ for every } j=\ell+1, \dots, n.
\tag*{}
\end{align}
We obtain the following:\footnote{If $\ell=0$, then the summation $\sum_{j=1}^{\ell} \alpha_j\cdot |(X-x)\cap \cf(E_{X-x}^j)|$ is defined to be $0$.}
\begin{align*}
&u(X-x)-u(X) \tag*{} \\
&=\Bigl\{\sum_{j=1}^{\ell} \alpha_j\cdot |(X-x)\cap \cf(E_{X-x}^j)|+\sum_{j=\ell+1}^{n}\alpha_j \cdot |(X-x)\cap \cf(E_{X-x}^j)|\Bigr\} \tag*{} \\
&-\Bigl\{\sum_{j=1}^{\ell}\alpha_j \cdot |X\cap \cf(E_{X}^j)|+\sum_{j=\ell+1}^{n}\alpha_j \cdot |X\cap \cf(E_{X}^j)| \Bigr\}
\tag*{} \\
&=\sum_{j=\ell+1}^{n}\alpha_j \cdot |(X-x)\cap \cf(E_{X-x}^j)|-\sum_{j=\ell+1}^{n}\alpha_j \cdot |X\cap \cf(E_{X}^j)| \tag*{} \\
&=0,
\end{align*} 
where
\begin{itemize}
\item the second equality follows from $X-x\in \Psi^j$ (which is implied by (\ref{11-22-1})), $X\in \Psi^j$ and $(b|j)$ (ii) for every $j$ with $1\leq j \leq \ell$, and
\item the third equality follows from (\ref{11-20-1}).
\end{itemize}
Hence,
\begin{align}
u(X-x)-u(X)=0 \label{11-26-2}
\end{align}
By $X-x\in \Psi^j$, $X\in \Psi^j$ and the first statement of $(b|j)$ for every $j$ with $1\leq j \leq \ell$, we have
\begin{align}
\delta_X^j=0 \text{ and } \delta_{X-x}^j=0 \text{ for every } j \text{ with } 1\leq j \leq \ell.
\label{12-2-1}
\end{align}
We obtain
\begin{align*}
\sum_{k=1}^n \delta_X^k\cdot |X\setminus C(E_X^k)|&=\delta^{\ell+1}_X\cdot |X\setminus C(E_X^{\ell+1})| \\
&>\delta^{\ell+1}_{X-x}\cdot |(X-x)\setminus C(E_{X-x}^{\ell+1})| \\
&=\sum_{k=1}^n \delta_{X-x}^k\cdot |(X-x)\setminus \cf(E_{X-x}^k)|,
\end{align*}
where the two equalities follow from (\ref{12-2-2}), (\ref{12-2-1}), and the definition of $\delta$, 
and the strict inequality follows from (\ref{11-20-0}) and (\ref{12-2-2}).
This inequality and (\ref{11-26-2}) imply
\begin{align*}
\tilde u(X-x)>\tilde u(X).
\end{align*}
Hence, condition (i) of ordinal concavity$^{++}$ holds for $x'=\emptyset$. 
\\
\textbf{Subcase 2-1-1-2:} Suppose $X\nsupseteq \cf(E_{\bar{X}}^{\ell+1})$. Since $\bar{X}=X\cup X'$ and $\bar{X}\supseteq C(E^{\ell+1}_{\bar{X}})$ (which follows from the assumption of Subcase 2-1-1), there exists $x'\in X'\setminus X$ such that $x'\in \cf(E_{\bar{X}}^{\ell+1})$. Then,
\begin{align}
&u(X-x+x')-u(X) \tag*{} \\
&=\Bigl\{\sum_{j=1}^{\ell+1} \alpha_j\cdot |(X-x+x')\cap \cf(E_{X-x+x'}^j)|+\sum_{j=\ell+2}^{n}\alpha_j \cdot |(X-x+x')\cap \cf(E_{X-x+x'}^j)|\Bigr\} \tag*{} \\
&-\Bigl\{\sum_{j=1}^{\ell+1}\alpha_j \cdot |X\cap \cf(E_{X}^j)|+\sum_{j=\ell+2}^{n}\alpha_j \cdot |X\cap \cf(E_{X}^j)| \Bigr\}
\tag*{} \\
&\geq \sum_{j=1}^{\ell+1} \alpha_j \cdot |(X-x+x')\cap \cf(E_{X-x+x'}^j)| \tag*{} \\
&-\Bigl\{\sum_{j=1}^{\ell+1}\alpha_j\cdot |X \cap \cf(E_{X}^j)|+\sum_{j=\ell+2}^{n} \alpha_j \cdot |X\cap \cf(E_{X}^j)|\Bigr\} \tag*{} \\
&=\alpha_{\ell+1} \cdot |(X-x+x')\cap \cf(E_{X-x+x'}^{\ell+1})| \tag*{} \\
&-\alpha_{\ell+1}\cdot |X \cap \cf(E_{X}^{\ell+1})|-\sum_{j=\ell+2}^{n} \alpha_j \cdot |X\cap \cf(E_{X}^j)| \tag*{} \\
&=\alpha_{\ell+1} \cdot |(X-x+x')\cap \cf(E_{\bar{X}}^{\ell+1})| \tag*{} \\
&-\alpha_{\ell+1}\cdot |X \cap \cf(E_{\bar{X}}^{\ell+1})|-\sum_{j=\ell+2}^{n} \alpha_j \cdot |X\cap \cf(E_{X}^j)| \tag*{} \\
&=\alpha_{\ell+1}-\sum_{j=\ell+2}^{n} \alpha_j \cdot |X\cap \cf(E_{X}^j)| \tag*{} \\
&\geq 1,  \tag*{}
\end{align}
where
\begin{itemize}
\item the first inequality follows from $\sum_{j=\ell+2}^{n}\alpha_j \cdot |(X-x+x')\cap \cf(E_{X-x+x'}^j)|\geq 0$,
\item the second equality follows from $X-x+x'\in \Psi^j$ (which is implied by (\ref{11-22-1})), $X\in \Psi^j$,
and $(b|j)$ (ii) for every $j$ with $1\leq j \leq \ell$,
\item the third equality follows from $X-x+x'\in \Psi^{\ell}$ (which is implied by (\ref{11-22-1})), $X\in \Psi^{\ell}$,
and $(b|\ell)$ (iii),\footnote{If $\ell=0$, then the third equality follows from $E_{X-x+x'}^1=E_X^1=E_{\bar{X}}^1=\mathcal{X}$.}
\item the fourth equality follows from the assumption of Subcase 2-1 and $x'\in \cf(E_{\bar{X}}^{\ell+1})$, and
\item the last inequality follows from the definition of $\alpha_{\ell+1}$.
\end{itemize}
It follows that $u(X-x+x')\geq u(X)+1$, which together with (\ref{tilde-strict}) implies $\tilde u(X-x+x')>\tilde u(X)$. Hence, condition (i) of ordinal concavity$^{++}$ holds. \\
\noindent
\textbf{Subcase 2-1-2} Suppose $k(X)<\ell$, which implies $X\notin \Psi^{k(X)+1}$. By (\ref{11-19-1-k}), we have $X\nsupseteq \bar{X}\cap \cf(E_{\bar{X}}^{k(X)+1})$.
By following the same line of the proof of Subcase 2-1-1-2 (with $k(X)$ playing the role of $\ell$),
there exists $x'\in X'\setminus X$ with $\tilde u(X-x+x')>\tilde u(X)$. Hence, condition (i) of ordinal concavity$^{++}$ holds.
\\
\textbf{Subcase 2-2:} Suppose that there exists $k\in \{k(X')+1, \dots, n\}$ such that $x\in \cf(E_{\bar{X}}^k)$. Together with
\begin{itemize}
\item $E_{X'+x}^{j}\subseteq E_{\bar{X}}^{j}$ for every $j=1, \dots, n$ (which follows from $X'+x\subseteq \bar{X}$ and Claim \ref{claim:representation}),
\item $x\in X'+x\subseteq E_{X'+x}^{j}$ (where the set-inclusion follows from the definition of $E$) 
for every $j=1, \dots, n$, and
\item the substitutes condition,
\end{itemize}
it implies that there exists $k'\in \{k(X')+1, \dots, n\}$ with $x\in \cf(E_{X'+x}^{k'})$. Let $k^*\in \{k(X')+1, \dots, n\}$ denote the minimum index such that $x\in \cf(E_{X'+x}^{k^*})$.

By (\ref{11-19-1-k}) and $X'\in \Psi^j$ for every $j$ with $1 \leq j \leq k(X')$,
\begin{align*}
X'+x\in \Psi^j \text{ for every } j \text{ with } 1\leq j\leq k(X').
\end{align*}
Together with $(b|j)$ (ii) for every $j$ with $1 \leq j \leq k(X')$, it implies
\begin{align}
(X'+x)\cap \cf(E_{X'+x}^j)=X' \cap \cf(E_{X'}^j) \text{ for every } j \text{ with } 1\leq j\leq k(X').
\label{11-22-10}
\end{align}
By $E^1_{X'}=E^1_{X'+x}=\mathcal{X}$ and $(b|j)$ (iii) for every $j$ with $1\leq j \leq k(X')$, we have
\begin{align}
E_{X'}^j=E_{X'+x}^j \text{ for every } j=1, \dots, k(X')+1.
\label{11-22-7}
\end{align}
If $k(X')+1<k^*$, then
\begin{align*}
E_{X'+x}^{k(X')+2}&=E_{X'+x}^{k(X')+1}\setminus \Bigl(\cf(E_{X'+x}^{k(X')+1})\setminus (X'+x)\Bigr) \\
&=E_{X'+x}^{k(X')+1}\setminus \Bigl(\cf(E_{X'+x}^{k(X')+1})\setminus X'\Bigr) \\
&=E_{X'}^{k(X')+1}\setminus \Bigl(\cf(E_{X'}^{k(X')+1})\setminus X'\Bigr) \\
&=E_{X'}^{k(X')+2},
\end{align*}
where the second equality follows from $x\notin \cf(E_{X'+x}^{k(X')+1})$ (which follows from $k(X')+1<k^*$ and the minimality of $k^*$) and the third equality follows from (\ref{11-22-7}).
If $k(X')+2<k^*$, then by the same argument as above, we obtain $E_{X'+x}^{k(X')+3}=E_{X'}^{k(X')+3}$.
Repeating this procedure, we obtain
\begin{align*}
E_{X'}^j=E_{X'+x}^j \text{ for every } j \text{ with } k(X')+2\leq j\leq k^*.
\end{align*}
This condition and (\ref{11-22-7}) imply
\begin{align}
E_{X'}^j=E_{X'+x}^j \text{ for every } j=1, \dots, k^*.
\label{11-26-1}
\end{align}
Together with the minimality of $k^*$, it yields
\begin{align}
(X'+x)\cap \cf(E_{X'+x}^j)=X' \cap \cf(E_{X'}^j)
\text{ for every } j \text{ with } k(X')+1\leq  j \leq  k^*-1.
\label{11-22-11}
\end{align}
We obtain
\begin{align}
&u(X'+x)-u(X') \tag*{} \\
&=\Bigl\{\sum_{j=1}^{k^*} \alpha_j\cdot |(X'+x)\cap \cf(E_{X'+x}^j)|+\sum_{j=k^*+1}^{n}\alpha_j \cdot |(X'+x)\cap \cf(E_{X'+x}^j)|\Bigr\} \tag*{} \\
&-\Bigl\{\sum_{j=1}^{k^*}\alpha_j \cdot |X'\cap \cf(E_{X'}^j)|+\sum_{j=k^*+1}^{n}\alpha_j \cdot |X'\cap \cf(E_{X'}^j)| \Bigr\}
\tag*{} \\
&\geq \sum_{j=1}^{k^*} \alpha_j \cdot |(X'+x)\cap \cf(E_{X'+x}^j)| \tag*{} \\
&-\Bigl\{\sum_{j=1}^{k^*}\alpha_j\cdot |X' \cap \cf(E_{X'}^j)|+\sum_{j=k^*+1}^{n} \alpha_j \cdot |X'\cap \cf(E_{X'}^j)|\Bigr\} \tag*{} \\
&=\alpha_{k^*} \cdot |(X'+x)\cap \cf(E_{X'+x}^{k^*})| \tag*{} \\
&-\alpha_{k^*}\cdot |X' \cap \cf(E_{X'}^{k^*})|-\sum_{j=k^*+1}^{n} \alpha_j \cdot |X'\cap \cf(E_{X'}^j)| \tag*{} \\
&=\alpha_{k^*}-\sum_{j=k^*+1}^{n} \alpha_j \cdot |X'\cap \cf(E_{X'}^j)| \tag*{} \\
&\geq 1,  \tag*{}
\end{align}
where
\begin{itemize}
\item the first inequality follows from $\sum_{j=k^*+1}^{n}\alpha_j \cdot |(X'+x)\cap \cf(E_{X'+x}^j)|\geq 0$,
\item the second equality follows from (\ref{11-22-10}) and (\ref{11-22-11}),
\item the third equality follows from (\ref{11-26-1}) and $x\in \cf(E_{X'+x}^{k^*})$, and
\item the last inequality follows from the definition of $\alpha_{k^*}$.
\end{itemize}
It follows that $u(X'+x)\geq u(X')+1$, which together with (\ref{tilde-strict}) implies $\tilde u(X'+x)>\tilde u(X')$. Hence, condition (ii) of ordinal concavity$^{++}$ holds.
\qed

\end{document}